\documentclass[qsecnum,amsmath,preprintnumbers,superscriptaddress,nofootinbib,aps,prd,10pt,a4paper]{revtex4-1}
\pdfoutput=1
\usepackage[utf8]{inputenc}
\usepackage[english]{babel}
\usepackage{amsmath}
\usepackage{amsfonts}
\usepackage{amssymb}
\usepackage{amsthm}
\usepackage{hyperref}
\usepackage{color}
\usepackage{tikz}
\usepackage{feynmp-auto}
\usepackage{bm}
\usepackage{graphicx}
\usetikzlibrary{matrix}

 \def\be{\begin{equation}}
\def\ee{\end{equation}}
 \def\ba{\begin{align}}
\def\ea{\end{align}}
\def\bea{\begin{eqnarray}}
\def\eea{\end{eqnarray}}

\def\a{\alpha}
\def\b{\beta}

\def\m{\mu}
\def\n{\nu}

\def\tg{\tilde{g}}

\def\tg{\tilde{g}}
\def\tn{\tilde{\nabla}}
\def\tR{\tilde{R}}

\begin{document}
\preprint{IFT-UAM/CSIC-20-85, FTUAM-20-10}
\title{{\bf Non-minimal Tinges of Unimodular Gravity}}
\author{Mario Herrero-Valea}
\email[]{mherrero@sissa.it}
\address{SISSA, Via Bonomea 265, 34136 Trieste, Italy and INFN Sezione di Trieste}
\address{IFPU - Institute for Fundamental Physics of the Universe \\Via Beirut 2, 34014 Trieste, Italy}

\author{Raquel Santos-Garcia}
\email[]{raquel.santosg@uam.es}
\address{Departamento de F\'isica Te\'orica and Instituto de F\'isica Te\'orica, IFT-UAM/CSIC,
Universidad Aut\'onoma de Madrid, Cantoblanco, 28049 Madrid, Spain}

\begin{abstract}
Unimodular Gravity is normally assumed to be equivalent to General Relativity for all matters but the character of the Cosmological Constant. Here we discuss this equivalence in the presence of a non-minimally coupled scalar field. We show that when we consider gravitation to be dynamical in a QFT sense, quantum corrections can distinguish both theories if the non-minimal coupling is non-vanishing. In order to show this, we construct a path integral formulation of Unimodular Gravity, fixing the complicated gauge invariance of the theory and computing all one-loop divergences. We find a combination of the couplings in the Lagrangian to which we can assign a physical meaning. It tells whether quantum gravitational phenomena can be ignored or not at a given energy scale. Its renormalization group flow differs depending on if it is computed in General Relativity or Unimodular Gravity.
\end{abstract}

\maketitle
\tableofcontents
\vspace{2cm}
\section{Introduction}\label{sec:introductions}
One of the most everlasting problems in theoretical physics is the Cosmological Constant problem \cite{Weinberg:1988cp,Martin:2012bt} -- the question of why our Universe is currently evolving according to the presence of a very small cosmological constant, corresponding to $M_P^2 \Lambda\sim 10^{-46}{\rm GeV}^4$, where $M_P\sim 10^{19} {\rm GeV}$ is the Planck mass. Being precise, this problem has actually two sides. The first belongs to the realm of model building and aims to describe which concrete physical mechanism leads to the observed value of $\Lambda $. Many attempts have been done in this direction during the last decades (see e.g. \cite{Gubitosi:2012hu, Joyce:2016vqv} and references therein), but so far we do not have any clear experimental signature that favours one or another. 

The second facet of the problem is of more fundamental theoretical nature. Even if a sensible mechanism to produce the current value of $\Lambda$ at the classical level is described, it still remains to explain why this value should be stable under radiative corrections. In a Quantum Field Theory (QFT), all dimensionful parameters receive corrections from loops of interacting particles that shift the classical value of parameters. This occurs even if the particles running in the loops do not manifest themselves in the low energy spectrum of the theory. In particular, if we think of an Effective Field Theory (EFT) setting, the cosmological constant receives contributions proportional to the cut-off of the theory, which encodes the ignorance about the UV degrees of freedom \cite{Cohen:2019wxr}. This means that in a gravitational theory described at low energies by General Relativity (GR), we expect corrections of the form $\delta(M_P^2 \Lambda)\sim M_P^4$, which are clearly much larger than the observed value of the cosmological constant. Although this \emph{hierarchy problem} can be solved by the inclusion of a very fine-tuned counter-term, it raises a question about the sensitivity of low energy observables to high energy degrees of freedom and thus poses a problem for the viability of the EFT, where separation of scales is critical. 

A possible way out of this issue is to modify the infra-red (IR) limit of the gravitational theory, so that the behavior of the cosmological constant gets replaced by a different dynamical avatar. This is the direction of research followed by massive gravity \cite{deRham:2014zqa,Dubovsky:2004sg}, where the graviton mass regulates the IR limit of GR; and of the plethora of (Beyond-) Hordensky/DHOST models \cite{Kobayashi:2019hrl}, where the dynamics of an extra scalar degree of freedom replaces the need for Dark Energy. However, the viability of both approaches has been recently questioned from different directions, and the allowed parameter space is shrinking quickly  \cite{Creminelli:2019kjy,Bellazzini:2017fep,Baker:2017hug,Ezquiaga:2017ekz,Creminelli:2017sry,Sakstein:2017xjx}. 

A particularly simple modification of GR that has attracted scattered attention during the last decades, although it is almost as old as GR itself\footnote{The equations of motion of UG appear for the first time ever in a 1919 paper by Einstein himself \cite{Einstein}. However, that work was not related to the cosmological constant but instead to the structure of point particles within GR.}, is Unimodular Gravity (UG) \cite{Unruh:1988in,Henneaux:1989zc,Kreuzer:1989ec}, formulated by appending the Einstein-Hilbert action with a condition of constant determinant for the metric tensor. Since the variation of this determinant is proportional to the trace of the equations of motion (eom), this effectively suppresses the trace degree of freedom of the metric. The resulting eom of UG are then the traceless part of Einstein equations \cite{Alvarez:2005iy,Ellis:2010uc,Buchmuller:1988wx}
\begin{align}\label{eq:einstein_traceless}
R_{\mu\nu}-\frac{1}{4}g_{\m\n}R=G\left(T_{\mu\nu}-\frac{1}{4}T g_{\m\nu}\right),
\end{align}
where $T_{\m\n}$ is the energy-momentum tensor of matter coupled to gravity. Any possible cosmological constant in the Lagrangian, or its radiative corrections, would be contained in the trace of Einstein equations and therefore they drop from the eom of UG.

Although this seems to signal a problem to reproduce well-known cosmological physics, it is not the case. The standard classical dynamics for gravity is recovered by the use of Bianchi identities, which are always true for a Riemannian manifold and imply, when taken together with \eqref{eq:einstein_traceless}
\begin{align}\label{eq:constraint}
\nabla^\m R_{\m\n}=\frac{1}{2}\nabla_\n R\rightarrow R+G T=4{\cal C},
\end{align}
after integration in a compact manifold without boundaries, and provided that $\nabla_\m T^{\m\n}=0$. Here ${\cal C}$ is an integration constant. If we now eliminate $T$ from \eqref{eq:einstein_traceless} by means of \eqref{eq:constraint} we recover the full set of Einstein equations
\begin{align}\label{eq:full_einstein}
R_{\m\n}-\frac{1}{2}R g_{\m\n} + {\cal C}g_{\m\n}=G T_{\m\n},
\end{align}
where ${\cal C}$ takes the role of a cosmological constant. However, here ${\cal C}$ is an integration constant instead of a coupling in the Lagrangian and therefore it does not receive radiative corrections \cite{Alvarez:2015sba}, effectively solving the second facet of the cosmological constant that we have discussed. The value of the cosmological constant is not given by vacuum energy but instead it is fixed by initial conditions when solving the eom. This mechanism has been explored in the context of inflation in \cite{Ellis:2013uxa}, while in \cite{Shaposhnikov:2008xb} it was exploited together with scale invariance to produce the complete thermal history of the Universe.
 
The fact that UG reproduces Einstein equations has led to a wide discussion of whether it is fully equivalent to GR -- apart form the discussed role of the cosmological constant-- or if there is some physical phenomenon that can serve to distinguish both theories (\cite{Alvarez:2016uog,Percacci:2017fsy,Saltas:2014cta} and references therein). From the previous discussion, it should be clear that (semi-)classically\footnote{By semi-classically here we mean quantum matter fields, represented by a quantum corrected energy-momentum tensor $\langle T_{\m\n}\rangle$, coupled to classical gravity by replacing $T_{\m\n}$ by $\langle T_{\m\n} \rangle$ in the equations of motion.} there cannot be any difference between both theories. The equations of motion are the same and the number of degrees of freedom propagated by UG matches those of GR -- a single massless graviton \cite{vanderBij:1981ym,Herrero-Valea:2018ilg}. The same must be true for any tree-level computation.

However, things are more subtle when dealing with the quantum nature of the gravitational field. In order to properly formulate a path integral for UG we need to resolve the constraint $|g|=1$ in an explicit way. As a consequence, and although on-shell states match those of GR, off-shell states are different, owing to a different gauge group. While the graviton fluctuation of GR is traceless only on the mass-shell, the one propagated by UG has a vanishing trace even for off-shell states. This means that loops with running gravitons are potentially different in both theories. 

Quantum phenomena in UG have been previosuly studied from several directions of research \cite{deBrito:2019umw,Eichhorn:2015bna,Saltas:2014cta,Smolin:2009ti,Alvarez:2016uog,Martin:2017ewb,Padilla:2014yea}. Of particular interest are \cite{Alvarez:2015sba,Ardon:2017atk}, where the one-loop effective action of UG is obtained by using two different approaches. Although the numerical results of both works differ, something which may be a gauge artefact, their physical conclusion is the same -- the cosmological constant does not renormalize and UG is one-loop finite. However, since in both works the theory is taken in vacuum, it is not possible to have access to any physical observable in order to compare the dynamics of UG with that of GR. This would require to couple another field to gravitation and account for its backreaction onto the geometry.

In this work we tackle this last point by considering UG together with the action for a non-minimally coupled scalar field. We will thus formulate a perturbative QFT expansion for UG coupled to matter, clarifying the issue of fixing the complicated gauge freedom of the theory and deriving all the elements required to implement perturbation theory around flat space. We will afterwards use these tools to compute the renormalization group (RG) flow of the different coupling constants in the action, at the one-loop level. This will allow us to identify a physically relevant essential coupling and compute its $\beta$-function, that we will be able to compare with the equivalent one as computed in GR.

This paper is organized as follows. First, in section \ref{sec:unimodular_gravity} we will describe Unimodular Gravity in more detail, together with the matter action that we consider. In order to quantize the system we will use the Background Field Method, described in section \ref{sec:bg_field} together with the concept of Weyl geometry and the BRST invariance of the gauge fixed action. We will later compute correlation functions at the one-loop order by expanding around flat space, as described in sections \ref{sec:perturbations} and \ref{sec:computation}, where we compute divergences in the $\overline{MS}$ scheme. Finally, we will derive the $\beta$-functions and anomalous dimensions of all the couplings in the one-loop effective action in section \ref{sec:beta_functions}, comparing our results with the general relativistic ones in \ref{sec:versus}. We will draw our conclusions in section \ref{sec:conclusions}. For completeness, we add two small appendices describing the computation in GR -- appendix \ref{sec:GR} -- and the discussion of divergences in UG in vacuum, in appendix \ref{sec:vacuum}.

\section{Unimodular Gravity}\label{sec:unimodular_gravity}
We \emph{define} UG by adding a condition of constant determinant to the Einstein-Hilbert action
\begin{align}
S=-\frac{1}{2G}\int d^4x \sqrt{|g|}\ R + S_{\rm matter},\quad |g|=\varepsilon,
\end{align} 
where $\varepsilon$ is a constant tensor density. In the following we will be interested on perturbations around flat space, so we fix it to $\varepsilon=1$ henceforth. Here $G=8 \pi M_P^{-2}$ is the Newton's constant.

As a consequence of the condition $|g|=1$, UG is not invariant under the full group of diffeomorphisms. Instead, it is invariant only under those that preserve the constraint, corresponding to \emph{volume preserving diffemorphisms, VDiff} \cite{Alvarez:2006uu,Gielen:2018pvk}. Their action is characterized at the infinitesimal level by a transverse vector
\begin{align}
\delta g_{\m\n}={\cal L}_{\xi}g_{\m\n},\quad \nabla_\m \xi^\m=0,
\end{align}
where ${\cal L}_\xi$ stands for the Lie derivative along $\xi^\mu$. We will dub the corresponding Lie algebra as \emph{TDiff} for this reason. In the rest of this document we will also use \emph{TDiff} in an sloppy way to refer to the full symmetry group. Effectively, we are replacing the four gauge constraints of GR by three of them -- those corresponding to the volume preserving subgroup-- plus the unit determinant constraint. The outcome for on-shell states is the same in both theories, four constraints that leave a single transverse and traceless graviton as the only propagating degree of freedom. However, this implies an important difference for off-shell states, since the constraint $|g|=1$ is also satisfied by them, unlike gauge constraints, which only act on physical degrees of freedom. As a consequence, the metric fluctuations of UG are always exactly traceless
\begin{align}\label{eq:traceless_variation}
\delta g_{\m\n}\equiv \delta \tilde{g}_{\m\n}-\frac{1}{4}\tilde{g}_{\m\n} \tilde{g}^{\a\b}\delta \tilde{g}_{\a\b},
\end{align} 
with $\tilde{g}_{\a\b}$ an unconstrained metric. Indeed, this is the reason as to why the eom are traceless, since they correspond to a variation with respect to this variable.

Although for classical matters we can use \eqref{eq:traceless_variation} to derive the eom, in order to perform a path integral over the gravitational field we need to resolve the constraint $|g|=1$. It must be included in the integration measure, giving
\begin{align}
{\cal Z}[T_{\m\n}]=\int [{\cal D} g]\ \delta\left(|g|-1\right) e^{iS+i T\cdot g},
\end{align}
where we have defined the dot product
\begin{align}
T\cdot g=\int d^4x\sqrt{|g|}\ T^{\m\n}g_{\m\n}.
\end{align}

Several ways to resolve this issue have been explored before, including using a Lagrange multiplier \cite{Buchmuller:1988wx} and a Stuckelberg field \cite{Henneaux:1989zc,Hammer:2020dqp,Jirousek:2018ago}. Here we choose to deal with it by performing a change of variables to a new metric defined by 
\begin{align}\label{eq:change_var}
g_{\m\n}=\tilde{g}_{\m\n} |\tilde{g}|^{\frac{1}{4}} ,
\end{align}
so that $|g|=1$ is satisfied identically. In terms of the new metric $\tilde{g}_{\m\n}$ and after integration by parts, the action of UG reads \cite{Alvarez:2012uz}
\begin{align}\label{eq:action}
S_{\rm UG}=-\frac{1}{2G}\int d^4x \ |\tilde g|^{\frac{1}{4}}\ \left(\tR + \frac{3}{32}\frac{\tn_\m |\tg| \tn^\m |\tg|}{|\tg|^2}\right),
\end{align}
where $\tg_{\m\n}$ is now an unconstrained field and variations can be taken freely.

Note that factors of $g$, which behaves as sort of an extra scalar field, have now appeared in the action. In a diffeomorphism invariant theory this is not possible, because the determinant of the metric transforms as a density under a general \emph{Diff} element. However, the fact that here we are dealing only with volume preserving transformations ensure that $g$ will transform as a true scalar and thus the new terms are allowed by symmetry.

The change of variables \eqref{eq:change_var} also introduces an extra \emph{fictitious} gauge symmetry in the form of Weyl invariance
\begin{align}\label{eq:Weyl}
\tg_{\m\n}\rightarrow \Omega(x)^2 \tg_{\m\n},
\end{align}
where $\Omega(x)$ is an arbitrary function of the space-time coordinates. The full gauge symmetry of the theory to be considered is then the direct product of \emph{TDiff} and \emph{Weyl}, a combination that has been dubbed \emph{WTDiff} before \cite{Alvarez:2006uu}. It is precisely Weyl invariance which comes to replace the determinant constraint in this form of the action, giving the extra condition needed to reduce the number of degrees of freedom to a single massless graviton.

Although we will use the action \eqref{eq:action} in order to evaluate the path integral of UG, we are interested on writing results in terms of the original metric variable, that we choose as our physical metric. This is achieved by simply choosing the gauge $|\tg|=1$ for Weyl transformations, thus identifying both metrics in \eqref{eq:change_var}. This is certainly true at the classical level, but one might be worried by the potential presence of a Weyl anomaly in the effective action, that would then obstruct the identification in the quantum variables. There are no reasons to worry, however, since it can be proven that the identification of the original metric $g_{\m\n}$ as the physical one precisely ensures the absence of anomalies \cite{Falls:2018olk,Percacci:2011uf}.

Note that when working with the action \eqref{eq:action} it is straightforward to understand the main feature of UG. Due to Weyl invariance, a cosmological constant term is forbidden in the action and it cannot be generated by radiative corrections either. Moreover, the Ward identity stemming from \eqref{eq:Weyl} precisely enforces the tracelessness of the eom and all subsequent variations
\begin{align}\label{eq:Ward_identity}
\tilde{g}^{\m_1\n_1}\tilde{g}^{\m_2\n_2}\dots \tilde{g}^{\m_n\n_n}\frac{\delta }{\delta \tilde{g}_{\m_1\n_1}}\frac{\delta }{\delta \tilde{g}_{\m_2\n_2}}\dots \frac{\delta S}{ \delta \tilde{g}_{\m_n\n_n}}=0.
\end{align}
This also implies that the graviton excitation $h_{\m\n}=\tg_{\m\n}-\eta_{\m\n}$ will always be exactly traceless. One can check explicitly that the eom derived from \eqref{eq:action}, in the gauge $|\tilde{g}|=1$ where we restore the original metric, are indeed the traceless Einstein equations \eqref{eq:einstein_traceless}.

As we discussed before, one of the main conundrums in the formulation of UG is the question of whether it is really a different theory than GR or if otherwise, and barring aside the role of the cosmological constant, they are exactly the same theory. Classically it is obvious that the answer is the latter. Since the eom of both theories are equivalent, the theories are so. However, quantum mechanically there are subtleties due to the different gauge group of UG. This question has been explored in several works from different points of view \cite{deBrito:2019umw,Eichhorn:2015bna,Saltas:2014cta,Smolin:2009ti,Alvarez:2016uog,Martin:2017ewb,Padilla:2014yea,Alvarez:2015sba,Ardon:2017atk}, but all of them consider the theory in vacuum, with only gravitation present. Although this is an interesting setting, the simplicity of the theory implies that nothing can be said about the true equivalence of the theories. In particular, if UG is considered alone, there are no physical observables that can be used to establish a comparison with GR, since the one-loop correction in vacuum is finite.

In order to bypass this problem and to be able to define dependable quantities to establish such a comparison, we couple here UG to matter. To keep things simpler -- but not trivial-- we will consider a toy model comprised of a single massive scalar field with a quartic interaction and non-minimal coupling to gravity
\begin{align}\label{eq:action_matter}
S_{\rm matter}=\int d^4x\ \left(\frac{1}{2}\partial_\m \phi \partial^\m \phi -\frac{m^2}{2}\phi^2 - \lambda\phi^4 -\frac{\xi}{2}\phi^2 R \right),
\end{align}
where we have already fixed $|g|=1$. Note that, as advertised, the action is written with respect to the original metric $g_{\m \n}$ that we consider as physical. 

An important difference with respect to GR arises here. Due to the constraint on the metric determinant, neither the mass term nor the quartic interaction couple directly to gravity. This will have a direct effect on the form of the vertices coupling gravity to matter in the perturbative expansion of this Lagrangian. Additionally, and since we will rely on perturbation theory for our later computations, we will always consider $G,\xi,\lambda \ll 1$.

Finally, note that since the Weyl invariance of the action \eqref{eq:action} appears here as a consequence of the change of variables \eqref{eq:change_var}, the scalar field \emph{is inert under it}. Unlike standard Weyl transformations, where a scalar would transform with a factor proportional to its energy dimension, here the symmetry is restricted purely to the metric sector. For this reason it has sometimes been dubbed as fake or spurious Weyl Invariance \cite{Percacci:2011uf}.

The total action that we will consider is then the sum of \eqref{eq:action} and \eqref{eq:action_matter}
\begin{align}
S=S_{\rm UG}+S_{\rm matter}.
\end{align}

However, in \eqref{eq:action_matter} the metric is unimodular. By performing the change of variables to the unconstrained metric $g_{\m\n}=|\tilde g|^{\frac{1}{4}}\tilde{g}_{\m\n}$ we have
\begin{align}\label{eq:final_action}
S=\int d^4x\bigg\{ &|\tilde g|^{\frac{1}{4}}\left[ -\frac{1}{2G}  \left(\tilde R + \frac{3}{32}\frac{\tilde \nabla_\m |\tilde g| \tilde \nabla^\m |\tilde g|}{|\tilde g|^2}\right) + \frac{1}{2}\partial_\m \phi \partial^\m \phi -\frac{\xi}{2}\phi^2 \left(\tilde R + \frac{3 \tilde 
\square |\tilde{g}|}{4|\tilde g|} - \frac{27 \tilde \nabla_{\m} |\tilde g| \tilde \nabla^{\m}|\tilde g|}{32 |\tilde g|^2}\right)\right]-\frac{m^2}{2}\phi^2 -\lambda \phi^4\bigg\},
\end{align}
where we have integrated by parts in some terms. All indices in this expression must be contracted by using the unconstrained metric $\tilde g_{\m\n}$. This is the action that we will use hereinafter.

\section{The Background Field Expansion}\label{sec:bg_field}
We will formulate the path integral of the theory by using standard tools. In order to be able to preserve explicitly the gauge invariance of gravitational correlation functions we will rely on the use of the background field method \cite{Abbott:1981ke,Barvinsky:2017zlx}. We thus start by defining the complete path integral that we will deal with as
\begin{align}
{\cal Z}[J_{\m\n}, j]=\int [{\cal D}\tilde{g}] [{\cal D}\phi]\ e^{i(S+ J\cdot \tilde{g} +  j\cdot \phi)},
\end{align}
where we have introduced two sources $J_{\m\n}$ and $j$, which couple to the metric and to the scalar field respectively. We will use those to define correlation functions in the usual way through variational derivatives with respect to them.

Now, following the background field method, we separate the metric into background and fluctuation by
\begin{align}
\tilde{g}_{\m\n}=\bar{g}_{\m\n}+h_{\m\n},
\end{align}
where $h_{\m\n}$ is the graviton field. Since this is just a shift of the integration variable, we can set $[{\cal D}\tilde{g}]=[{\cal D}h]$. 

Under this redefinition, the exponent inside the path integral can be expanded in powers of $h_{\m\n}$
\begin{align}
S_J=\left. S_J \right|_{\tilde{g}=\bar{g}}+ \int d^4 x\ \left.\frac{\delta S_J}{\delta \tilde{g}(x)_{\m\n}}\right|_{\tilde{g} = \bar g}h(x)_{\m\n}+\frac{1}{2}\int d^4x \int d^4y \ h(x)_{\m\n}\left.\frac{\delta^2  S_J}{\delta \tilde{g}(x)_{\m\n}\delta \tilde{g}(y)_{\a\b}}\right|_{\tilde{g} = \bar g}h(y)_{\a\b}+{\cal O}(h^3),
\end{align}
where we have defined $S_J=S+J\cdot \tilde{g} +  j\cdot \phi$. The first term in the expansion corresponds to the action evaluated in the background field, while the linear term vanishes whenever the background configuration satisfies the classical equations of motion. Since in this work we are interested only in one-loop effects, we cut the expansion at second order, which corresponds to leading order in the $\hbar$ expansion.

Since we have shifted the integration variable to $h_{\m\n}$, the background metric can be thought as an extra source, with the path integral depending on it
\begin{align}
{\cal Z}[J_{\m\n},\bar{g}_{\m\n},j]=\int [{\cal D}h] [{\cal D}\phi]\ e^{i S_J}.
\end{align}

If we now define the Quantum Effective Action in the standard way by a Legendre transform before and after the field redefinition, we find the apparently trivial identity
\begin{align}
\Gamma[\tilde{g}_{\m\n},\phi]=\Gamma[\bar{g}_{\m\n}+h_{\m\n},\phi].
\end{align}

However, this is not trivial at all. It means that, due to the appearance of the background metric as a shift of the total one, we can capture any covariant term of the Quantum Effective Action just by computing those correlators in which only $\bar{g}_{\m\n}$ and $\phi$ appear on external legs, while $h_{\m\n}$ is a pure internal variable over which we integrate. This will clearly make our lives easier and defines our computational strategy.

The other advantage of the background field method is that it allows us to preserve the gauge invariance -- \emph{WTDiff} in this case -- of the Quantum Effective Action easily, by preserving that of any operator involving the background metric. This is due to the fact that, after the field redefinition, infinitesimal gauge transformations can be split in two
\begin{align}
&\delta_{\rm bg}\bar g_{\m\n}={\cal L}_{ \xi}\bar g_{\m\n}+ 2\omega \bar g_{\m\n},\label{eq:bg_1}\\
&\delta_{\rm bg}h_{\m\n}={\cal L}_{ \xi}h_{\m\n}+ 2\omega h_{\m\n},\label{eq:bg_2}\\
&\delta_{\rm q} \bar g_{\m\n}=0,\label{eq:q_1}\\
&\delta_{\rm q}h_{\m\n}={\cal L}_{ \xi}(\bar{g}_{\m\n}+h_{\m\n})+ 2\omega (\bar{g}_{\m\n}+h_{\m\n}),\label{eq:q_2}
\end{align}
where $\omega$ is the infinitesimal parameter associated to Weyl transformations \eqref{eq:Weyl} $\Omega(x)=1+\omega+{\cal O}(\omega^2)$.

As we see, $\delta_{\rm bg}$ corresponds to the gauge invariance of the background quantities, where $h_{\m\n}$ is then regarded as a tensor transforming in the same way as the metric. To this we must append the condition that $\phi$ is a scalar field inert under Weyl transformations.

Since the path integral that we need to compute integrates only over $h_{\m\n}$ and $\phi$, while $\bar{g}_{\m\n}$ is regarded as a source, we will only need to gauge fix the quantum part of the symmetry $\delta_{\rm q}$. The background symmetry will remain unaltered and therefore gauge invariance of our results is automatically ensured. All correlation functions must then satisfy an analogous expression to \eqref{eq:Ward_identity}, with the classical action $S$ replaced by the Quantum Effective Action $\Gamma[\tilde{g}_{\m\n},\phi]$.

\subsection{Weyl Geometry}

We thus turn now our attention to the issue of gauge fixing $\delta_{\rm q}$. In order to do that, we first wish to be able to construct a gauge fixing term which is invariant under the background remaining \emph{WTDiff} symmetry represented by $\delta_{\rm bg}$. A priori this does not seem like a complicated task, but the complexity of the gauge sector of the theory (cf. later) can make it a cumbersome task. In order to make things easier and more straightforward, we will use here the formalism introduced in \cite{Codello:2012sn,Iorio:1996ad} and named as \emph{Weyl Geometry}. By defining a full geometric construction which is explicitly Weyl covariant -- as well as diffeomorphism covariant-- we can construct invariant quantities in a easy way. 

The core of the method consists in the introduction of a $U(1)$ gauge field $W_\m$ which will serve to define Weyl covariant derivatives. However, since this is a Weyl invariant theory, this field \emph{is not} an external ingredient, but instead it can be built out of the fields already in the action. In our case, we define it to be
\begin{align}
W_\mu = \frac{1}{8} \bar{\nabla}_\m \log(|\bar{g}|).
\end{align}  
It can be easily checked that under a Weyl transformation \eqref{eq:Weyl}, $W_\mu$ behaves indeed as a $U(1)$ gauge field 
\begin{align}
W_\m\rightarrow W_{\m}+ \Omega \bar \nabla_\mu \Omega.
\end{align}

Using it we introduce a non-metric connection 
\begin{align}
\Gamma^{{\rm (W)}\a}_{\m\n}=\left\{\ \right\}^\a_{\m\n}-\delta^\a_\m W_\n -\delta^\a_\n W_\m + \bar g_{\m\n}W^\a,
\end{align}
where $\left\{\ \right\}^\a_{\m\n}$ is the Levi-Civita connection of the metric $\bar{g}_{\m\n}$. 
As usual, the connection $\Gamma^{{\rm (W)}\a}$ will induce a covariant derivative, that we label $\nabla^{\rm (W)}$. 

We complete the construction presented here by introducing the Weyl covariant derivative acting on a generic tensor ${\cal T}$
\begin{align}
D_\m{\cal T}=\nabla_\m^{(W)}{\cal T}-\lambda_{\cal T} {\cal T},
\end{align}
where $\lambda_{\cal T}$ is the scaling dimension of the tensor, defined as the weight of $\Omega$ under a Weyl transformation
\begin{align}
{\cal T}\rightarrow \Omega^{\lambda_{\cal T}}{\cal T}.
\end{align}
Note that, when defined in this way, $D_\m$ is compatible with the background metric $D_\m \bar{g}_{\a\b}=0$.

For the future it will be also useful to define a Weyl covariant curvature by using the Ricci identity acting on a generic vector $V^\a$
\begin{align}
[D_\m,D_\n]V^\a={\cal R}_{\m\n\ \ \b}^{ \ \ \ \a }V^\b,
\end{align}
which gives
\begin{align}
\nonumber {\cal R}_{\m\n\a\b}&=\bar{R}_{\m\n\a\b}+\bar{g}_{\m\a}\left(\bar{\nabla}_\n W_\b + W_\n W_\b\right)-\bar{g}_{\m\b}\left(\bar{\nabla}_\n W_\a + W_\n W_\a\right)-\bar{g}_{\n\a}\left(\bar{\nabla}_\m W_\b + W_\m W_\b\right)\\
&+\bar{g}_{\n\b}\left(\bar{\nabla}_\m W_\a + W_\m W_\a\right)-(\bar{g}_{\m\a}\bar{g}_{\n\b}-\bar{g}_{\m\b}\bar{g}_{\n\a})W^2,
\end{align}
and subsequently
\begin{align}\label{eq:ricci_tensor_D}
&{\cal R}_{\m\n}=\bar{R}_{\m\n}+2 W_\m W_\n + \bar{\nabla}_\m W_\n + \bar{\nabla}_\n W_\m -2 \bar{g}_{\m\n}W^2 + \bar{g}_{\m\n}\bar{\nabla}^\a W_\a, \\
&{\cal R}=\bar{R}+6 (\bar{\nabla}^\m W_\m -  W^2),
\end{align}
where $\bar{R}_{\m\n\a\b}$ is the Riemann tensor of the background metric.

The advantage of using $D_\m$ now is clear. For any tensor ${\cal T}$ with a well-defined scaling dimension -- that is, that there are not derivatives of $\Omega$ involved in the transformation of the tensor--, $D_\m {\cal T}$ will transform in the same way as ${\cal T}$. Constructing Weyl invariant quantities is just a matter of combining $D_\m$ with powers of $|\bar g|$ -- which enjoys scaling dimension $\lambda_{|\bar g|}=8$ -- to form scalars under Weyl transformation. A simple example of this is the action \eqref{eq:action} evaluated in $\bar{g}$, which can be easily written as
\begin{align}
\left.S\right|_{\tilde{g}=\bar{g}}=-\frac{1}{2G}\int d^4 x \ |\bar{g}|^{\frac{1}{4}}{\cal R}.
\end{align}

\subsection{Gauge Fixing and BRST invariance}
We finally turn ourselves to the problem of fixing the \emph{WTDiff} symmetry of the fluctuations. In principle one could think on attempting to fix the symmetry in a standard manner, by introducing a gauge fixing condition 
\begin{align}
F_\m=0,
\end{align}
and appending the action with a gauge fixing term
\begin{align}\label{eq:gauge_fixing_bad}
S_{\rm gf}=\int d^4x \ F_\m F^\m,
\end{align}
and the corresponding action for the ghosts. However, this is not as straightforward as it seems for two reasons. First, we are dealing with the direct product of \emph{TDiff} and \emph{Weyl}. The total number of conditions required to fix the symmetry is still four and thus it seems that choosing a space-time vector $F_\m$ does the work. However, the resulting gauge fixing term must then satisfy three conditions
\begin{enumerate}
\item It must break the quantum part of \emph{TDiff} invariance.
\item It must break the quantum part of \emph{Weyl} invariance.
\item It must preserve the background \emph{WTDiff} symmetry.
\end{enumerate}

As far as we know, it is not possible to choose a function $F_\m$ such that \eqref{eq:gauge_fixing_bad} satisfies all three conditions.

The other reason as to why a standard gauge fixing method is cumbersome is related to the structure of the \emph{TDiff} group. Since the generator of transverse diffeomorphisms is constrained to be transverse, the same will be true for the corresponding ghost field $c^\m$
\begin{align}\label{eq:transverse_c}
\bar \nabla_\m c^\m=0.
\end{align}

This condition has to be included in the measure in some manner. The easiest way is to follow \cite{Alvarez:2015sba} and use a transverse projector to project an otherwise arbitrary field
\begin{align}
c^\m=\left(\delta^\m_\n-\bar \nabla^\m\square^{-1} \bar \nabla_\n\right)d^\m,
\end{align}
where the inverse of the Laplace operator $\square=\bar \nabla_\m \bar{\nabla}^\m$ is defined by acting on an arbitrary tensor
\begin{align}
\square^{-1}\ \square =\square\  \square^{-1} {\cal T}={\cal T}.
\end{align}

By doing this, in a similar way to what happens with the transformation of the metric \eqref{eq:change_var}, we replace the condition \eqref{eq:transverse_c} by a $U(1)$ gauge symmetry acting on $d^\m$
\begin{align}\label{eq:project_1}
d^\m\rightarrow \bar{\nabla}^\m f,
\end{align}
where $f$ is an arbitrary function of the space-time coordinates. Thus, we will need to introduce a gauge fixing term for this symmetry as well, that will then generate a full new set of ghosts and anti-ghosts, of bosonic character this time. These have been sometimes dubbed in the literature as Nielsen-Kallosh ghosts \cite{Nielsen:1978mp,Kallosh:1978de}.

In order to circumvent all the complications implied by these properties, we decide here to fix the gauge by using BRST invariance \cite{Becchi:1975nq}. We thus introduce an operator $\mathfrak{s}$ which, when acting on the graviton fluctuation $h_{\m\n}$, implements a gauge transformation with the infinitesimal generator replaced by a ghost
\begin{align}\label{eq:BRST_h}
\mathfrak{s} h_{\m\n}={\cal L}_{c}(\bar{g}_{\m\n}+h_{\m\n})+2b ( \bar g_{\m\n}+ h_{\m\n}),
\end{align}
where we have introduced the ghost field associated to Weyl invariance $b$. Note that since the generators of the transformation are now Grassman variables, the operator $\mathfrak{s}$ is Grassman odd. To this we must append the transformation rules for the ghost fields $c^\m$ and $b$
\begin{align}
&\mathfrak{s}c^\m={\cal L}_{c}c^\m =c^\rho \nabla_\rho c^\m,\\
&\mathfrak{s}b={\cal L}_{c}b=c^\rho \nabla_\rho b,
\end{align}
which are inert under Weyl transformations.

However, in this work we are only interested in one-loop corrections, that we have already established that correspond to the quadratic approximation in the path integral. Since, as we will see later, the transformation of the ghost will always come in the final gauge-fixed action multiplied by another quantum field, we can just neglect the transformation of both ghost fields and write instead
\begin{align}
&\mathfrak{s}c^\m = {\cal O}({\rm field}^2) ,\\
&\mathfrak{s}b={\cal O}({\rm field}^2).
\end{align}

As before, the ghost field $c^\mu$ is forced to satisfy a transversality condition. However, since our ultimate goal is to obtain a gauge fixing term which preserves background \emph{WTDiff} invariance, from now on we define the transverse condition by using the Weyl covariant derivative
\begin{align}\label{eq:ghost_transverse_W}
D_\m c^\m=0.
\end{align}
We will do the same in any other BRST transformation from now on.

Note that this replacement is always possible, since we can always use whatever derivative we desire to compute the Lie derivative in \eqref{eq:BRST_h}. Alternatively, any difference between derivatives can be also absorbed in a redefinition of the ghost field $b$.

Again, we use a transverse projector to satisfy \eqref{eq:ghost_transverse_W}, which in this case will be given by
\begin{align}\label{eq:change_var_ghosts}
c^\m=\left(\delta^\m_\n -D^\m (D^2)^{-1} D_\n\right)d^\n,
\end{align}
and the inherited $U(1)$ invariance will take the form
\begin{align}
d^\m \rightarrow D^\m f.
\end{align}

In general, dealing with this kind of open algebra would require the sophisticated technique of BV quantization \cite{Batalin:1984jr}. However, in the case of UG things are simple enough so that we can construct the gauge fixing sector by simply including the gauge symmetry of the ghost field in the BRST operator \cite{Alvarez:2015sba,Baulieu:2020obv,Upadhyay:2015fna}. Consequently, we extend the action of $\mathfrak{s}$ appropriately. We introduce a ghost field $\alpha$ and write
\begin{align}
\mathfrak{s}d^\m=D^\m \alpha,
\end{align}
where, due to the Grassman parity of $\mathfrak{s}$, we see that $\alpha$ must be a \emph{bosonic} field.

We now append our theory with a complementary set of anti-ghost and auxiliary fields with the goal of closing the algebra of the BRST operator, that we demand to be nilpotent when acting on any field involved in the path integral
\begin{align}
\mathfrak{s}^2=0.
\end{align}
For symmetries whose associated ghost is Grassman odd, it is enough to add a single anti-ghost and an auxiliary field with even Grassman number to achieve the closure of the algebra. However, for symmetries whose ghost is bosonic, such as $\alpha$, things are more subtle. Since the auxiliary field needs to be Grassman odd, it is impossible to form its square. In that case we are required to introduce two pairs of anti-ghost and auxiliary fields. Following these rules and taking into account that here we have three gauge symmetries -- \emph{TDiff}, \emph{Weyl} and the $U(1)$ symmetry of the ghost field -- we find that we need the following set of fields and transformations
\begin{align}
&\mathfrak{s}\bar{c}^\mu =\rho^\mu,\quad \mathfrak{s}\rho^\mu=0,\\
&\mathfrak{s}\bar{b} =l,\quad \mathfrak{s}l=0,\\
&\mathfrak{s}x=m,\quad \mathfrak{s}m=0,\\
&\mathfrak{s}\bar{x}=\bar{m},\quad \mathfrak{s}\bar{m}=0.
\end{align}

Here the first two lines correspond to the auxiliary fields needed to close the algebra for \emph{WTDiff} transformations, while the rest are the two pairs of field required for the $U(1)$. Their Grassman character is
\begin{align}
&\{\bar{c}^\m, \bar{b},m,\bar{m}\}\equiv \text{Grassman odd}.\\
&\{\rho^\m, l, x, \bar{x}\}\equiv \text{Grassman even}.
\end{align}
This is enough to ensure the nil-potency of the BRST operator when acting on any field of the path integral within the one-loop approximation.

Once the action of $\mathfrak{s}$ onto every field is defined, we introduce the BRST gauge-fixing term, which includes the action of the ghosts, as the result of acting with $\mathfrak{s}$ on a so-called \emph{gauge fermion}
\begin{align}
S_{\rm BRST}=-\frac{1}{2G}\int d^4x\ \mathfrak{s}\Psi,
\end{align}
where $\Psi$ is a term quadratic in the fields and of odd Grassman parity. Thanks to the nil-potency of $\mathfrak{s}$ and once $S_{\rm BRST}$ is chosen in this way, the total action is invariant under a BRST transformation. The associated Ward-Takahashi identities then become the Slavnov-Taylor identities of the theory, ensuring a successful quantization.

The construction of $\Psi$ now replaces the arbitrary choice of gauge function $F^\m$. As long as $\Psi$ is Grassman odd and breaks gauge invariance -- but not BRST invariance -- it is a valid choice. Here we will however follow a conservative approach, still defining a gauge condition
\begin{align}\label{eq:gauge_condition}
F_\mu=D^\n h_{\m\n}+\tau D_\m h,
\end{align}
and writing
\begin{align}
\Psi=|\bar{g}|^{\frac{1}{4}}(\bar{c}_\m + D_\m \bar b)\left(  F^\m -\frac{1}{4\sigma}(\rho^\m - D^\m l)\right)+x\left(D_\m d^\m +\frac{1}{2\gamma}\bar{m}\right)+y\  \bar{x}\left(\bar g^{\frac{1}{4}}D_\m \bar{c}^\m -\frac{1}{2\gamma}m\right).
\end{align}
Here $\sigma$, $\tau$, $y$ and $\gamma$ are gauge parameters whose value we can use either to simplify our computations or to test gauge invariance of our results. The powers of $|g|$ are chosen so that the expression is invariant under background Weyl invariance. The form of this gauge fermion is motivated by the BRST formulation of the usual Faddev-Poppov gauge fixing method. If we were dealing with a simpler symmetry, and in the absence of Weyl invariance, then the first term would be enough to fix it and after integration of the auxiliary field $\rho$ we would have recovered the standard gauge fixing plus ghost action. Here the first term deals with the combined \emph{WTDiff} gauge symmetry while the rest is needed to be able to fix the $U(1)$ symmetry of the ghost sector.

Acting with the BRST operator we then have
\begin{align}
\nonumber S_{\rm BRST}=-\frac{1}{2G}\int d^4 x\, \bigg\{ &|\bar{g}|^{\frac{1}{4}}(\rho_\m +D_\m l)\left(F^\m -\frac{1}{4\sigma}(\rho^\m -D^\m l)\right)+|\bar{g}|^{\frac{1}{4}}(\bar{c}_\m+D_\m \bar b) \ \mathfrak{s}F^\m\\
&+ m(D_\m d^\m +\frac{1}{2\gamma}\bar{m})+|\bar g|^{\frac{1}{4}}x D^2 \alpha + y \ \bar{m}(|\bar{g}|^{\frac{1}{4}}D_\m \bar{c}^\m - \frac{1}{2\gamma}m)+y|\bar{g}|^{\frac{1}{4}}\ \bar{x}D_\m \rho^\mu\bigg\}.
\end{align}

Examining this expression we see that $\rho^\mu$, $m$ and $\bar{m}$ are linearly coupled, entering the path integral as sources. We can thus integrate them out by using their equations of motion. This simplifies the BRST term to
\begin{align}
S_{\rm BRST}=-\frac{1}{2G}\int d^4 x\,|\bar{g}|^{\frac{1}{4}} \bigg\{ &(\bar{c}_\m + D_\m \bar{b})\  \mathfrak{s}F^\m + \frac{2\gamma y}{y+1}\ D_\m \bar{c}^\m D_\n d^\n+ \sigma\left( F_\m -y D_\m \bar{x}\right)^2 +\frac{1}{4\sigma}  D_\m lD^\m l +D_\m l F^\m+  xD^2 \alpha\bigg\}.
\end{align}

Finally, appending this action to the classical action, we can write the path integral of UG in the background field approach to be
\begin{align}\label{eq:path_integral_final}
{\cal Z}[J_{\m\n},\bar{g}_{\m\n},j]=\int [{\cal D}h][{\cal D}\phi][{\cal D}\bar{c}][{\cal D}d][{\cal D}\bar{b}][{\cal D}b][{\cal D}\bar{x}][{\cal D}l][{\cal D}x][{\cal D}\alpha]\ e^{i (S_J+S_{\rm BRST})}.
\end{align}

\section{Perturbations around Flat Space}\label{sec:perturbations}
Once the path integral for the unimodular scalar-tensor theory is properly defined, we come to the task of computing the one-loop correction to the coupling constants. Since background \emph{WTDiff} invariance is ensured by construction, we will perform our computation by expanding the background metric around flat space-time
\begin{align}
\bar{g}_{\m\n}=\eta_{\m\n}+H_{\m\n},
\end{align}
where we will dub $H_{\m\n}$ as the \emph{background graviton fluctuation}. This will allow us to use standard techniques to compute Feynman diagrams. Correlation functions of the background metric will become correlators of $H_{\m\n}$ and we will capture the renormalization of the coupling constants by computing diagrams with $H_{\m\n}$ and $\phi$ in the external legs.

\subsection{Propagators for Bosonic Fields}
We start by computing the propagators of the fluctuations. In order to do that we take $S_{J}+S_{\rm BRST}$ and we set the background metric to be flat, thus retaining only the terms quadratic in the quantum fields. The lagrangian for the bosonic fields then reads
\begin{align}\label{eq:S2}
\nonumber {\cal L}_{2}=&-\frac{1}{2G}\left[h^{\m\n}\partial_\a \partial_\n h^\a_\m+h^{\a\b}\partial_\b \partial_\m h^\m_\a+\frac{1}{2}\partial^\b h^{\a\m} \partial_\m h_{\a\b}-\frac{1}{16}h \partial^2 h-\frac{1}{4}h^{\a\b}\partial^2 h_{\a\b}-\frac{1}{4}h \partial_\a \partial_\b h^{\a\b}-h^{\a\b}\partial_\a\partial_\b h\right.\\
\nonumber & +(\sigma-1) \partial_\a h^{\a\b}\partial_\m h^\m_\a -\frac{1+8 \sigma \tau}{4}\partial_\a h^\b_\a \partial^\a h-\frac{1+32 \sigma \tau^2}{32}\partial_\a h \partial^\a h+\frac{1}{4\sigma}\partial_\m l \partial^\m l + \sigma y^2 \partial_\m \bar{x}\partial^\m \bar{x}+ \partial_\a l (\partial_\b h^{\a\b}+\tau \partial^a h)\\
&\left. -2\sigma y \partial_\a \bar{x} (\partial_\b h^{\a\b}+\tau \partial^a h)+x\partial^2 \alpha\right]+\frac{1}{2}\left(\partial_\a \phi \partial^\a \phi -m^2 \phi^2\right).
\end{align}
We leave the discussion of the ghost sector involving $d^\m$, $\bar{c}^\m$, $b$ and $\bar{b}$ for the next subsection.

Here we find a striking difference between GR and UG. Due to the complicated gauge fixing sector involving bosonic Nielsen-Kallosh ghost fields, we find that the graviton fluctuation $h_{\m\n}$ mixes with the bosonic ghosts at the kinetic level, as indicated by the last terms in the second line in \eqref{eq:S2}. This means that in order to compute the propagator of the gravitational field, we need to take these fields into account in order to cancel spurious gauge pole contributions. It is not enough to take the $F_\m F^\m$ term in the gauge fixing and invert the kinetic term for the graviton by itself, \emph{even for tree-level computations}.

We take the action \eqref{eq:S2}, Fourier transforming it to momentum space and we write it in matrix form 
\begin{align}
{\cal L}_2=\frac{1}{2} \begin{pmatrix}
h_{\m\n},\bar{x},
l,
\phi,
x,
\alpha
\end{pmatrix} {\cal M}^{-1}(q)\begin{pmatrix}
h_{\a\b}\\
\bar{x}\\
l\\
\phi\\
x\\
\alpha
\end{pmatrix}
\end{align}
where ${\cal M}^{-1}(q)$ is the matrix-valued inverse propagator. Inverting it with the following sign convention 
\begin{align}
{\cal M}^{-1}(q){\cal M}(q)=i\mathbb{I},
\end{align}
gives the following non-vanishing propagators for the fields
\begin{align}\label{eq:proph}
\nonumber \langle h_{\m\n}(-q)h_{\a\b}(q)\rangle&=\frac{2iG}{q^2}\left(\eta_{\m\a}\eta_{\n\b}+\eta_{\m\b}\eta_{\n\a}-\frac{1+2\sigma(3+8\tau (1+\tau))}{\sigma (1+4\tau)^2} \eta_{\m\n}\eta_{\a\b}\right)+\frac{4iG}{q^6}\frac{1-2\sigma}{\sigma}q_\m q_\n q_\a q_\b\\
&+\frac{4iG}{q^4}\frac{3+4\tau}{1+4\tau}\left(\eta_{\m\n}q_\a q_\b+\eta_{\a\b} q_\m q_\n \right)-\frac{iG}{q^4}\frac{1+2\sigma}{\sigma}\left(\eta_{\m\a}q_\n q_\b +\eta_{\n\a}q_\m q_\b +\eta_{\m\b}q_\n q_\a +\eta_{\n\b}q_\m q_\a\right),
\end{align}

\begin{align}
&\langle l(-q)l(q)\rangle =-6i G \sigma \frac{1}{q^2},\\
&\langle l(-q) h_{\m\n}(q)\rangle = \frac{2 i G }{(1+4 \tau)}\frac{\eta_{\m\n}}{q^2},\\
&\langle \bar{x}(-q)\bar{x}(q) \rangle=-\frac{3 i G}{2\sigma y^2}\frac{1}{q^2},\\
&\langle \bar{x}(-q) h_{\m\n}(q)\rangle=-\frac{iG}{\sigma y (1+4\tau))}\frac{\eta_{\m\n}}{q^2}, \\
&\langle l(-q) \bar{x}(q) \rangle=\frac{iG}{y}\frac{1}{q^2},\\
&\langle x(-q) \alpha(q) \rangle=\frac{4 i G}{q^2},\\
&\langle \phi(-q)\phi(q)\rangle = \frac{i}{q^2-m^2}\label{eq:prop_phi}.
\end{align}

In order to simplify our computations we will set the gauge parameter $\tau=-3/4$, for which the graviton propagator reduces to
\begin{align}
\nonumber \langle h_{\m\n}(-q)h_{\a\b}(q)\rangle&=\frac{2iG}{q^2}\left(\eta_{\m\a}\eta_{\n\b}+\eta_{\m\b}\eta_{\n\a}-\frac{1+3\sigma}{4\sigma} \eta_{\m\n}\eta_{\a\b}\right)+\frac{4iG}{q^6}\frac{1-2\sigma}{\sigma}q_\m q_\n q_\a q_\b\\
&-\frac{iG}{q^4}\frac{1+2\sigma}{\sigma}\left(\eta_{\m\a}q_\n q_\b +\eta_{\n\a}q_\m q_\b +\eta_{\m\b}q_\n q_\a +\eta_{\n\b}q_\m q_\a\right).
\end{align}
In principle we could further simplify this expression by choosing $\sigma=-1/2$. However, we refrain to do so in order to be able to track the gauge dependence of our results along the computation. We will also leave the parameter $y$ arbitrary.

\subsection{The ghost Propagators}
We now focus in the action for the ghost fields
\begin{align}
S_{\rm gh}=-\frac{1}{2G}\int d^4 x\,|\bar{g}|^{\frac{1}{4}}\left[(\bar{c}_\m + D_\m \bar{b})\mathfrak{s}F^\m+ \frac{2\gamma y}{y+1}\ D_\m \bar{c}^\m D_\n d^\n\right],
\end{align}
with the goal of computing their propagators. 

Acting with the BRST operator on $F^\m$ gives
\begin{align}
&\mathfrak{s}F^\m=D^2 c^\m +{\cal R}^\m_\n c^\n +(2+8 \tau)D^\m b,
\end{align}
with ${\cal R}_{\m\n}$ given by \eqref{eq:ricci_tensor_D}. However, this is written in terms of the constrained field $c^\m$. We thus perform the change of variables \eqref{eq:change_var_ghosts} and write
\begin{align}
\mathfrak{s}F^\m=D^2 d^\m -D^\m D_\n d^\n+{\cal R}^\m_\n d^\n -2{\cal R}^\m_\n D^\n\left(D^{2}\right)^{-1}D_\a d^\a+(2+8 \tau)D^\m b.
\end{align}

Setting the background metric to be flat in order to derive the propagator, we have $D_\m\equiv \partial_\m$ and therefore the non-local operator $(\partial^2)^{-1}$ has a well-defined representation in momentum space when acting on an arbitrary tensor, given by
\begin{align}
(\partial^2)^{-1}{\cal T}=\int \frac{d^4 q}{2\pi} \left(-\frac{1}{q^2}\right){\cal T}\ e^{i q\cdot x}.
\end{align}
However, this will never enter into the definition of the propagators, since it comes multiplied by a curvature, which vanishes when $\bar{g}_{\m\n}=\eta_{\m\n}$. It will be important later when deriving the interaction vertices, but due to the same reason and since we will only need vertices with one external graviton, we will never have to workout the task of inverting this in general, but only take its flat realization.

Around flat space-time, the action that defines the propagator then gives the following Lagrangian
\begin{align}
{\cal L}_2^{\rm (gh)}=\left(\bar{c}_{\m} + \partial_\m \bar{b} \right)\left(\partial^2 d^\m -\partial^\m \partial_\n d^\n+(2+8 \tau)\partial^\m b\right)-z \ \partial_\m \bar{c}^\m \partial_\n d^\n,
\end{align}
where we must note that the different ghost sectors, belonging to \emph{TDiff} and \emph{Weyl}, are mixed at the kinetic level. Here we have defined $z=-2\gamma y (1+y)^{-1}$.

As with the bosonic fields, we now write this in matrix form after integration by parts
\begin{align}
{\cal L}_2^{\rm (gh)}=\begin{pmatrix}
\bar{c}_{\m},\bar{b}
\end{pmatrix} {\cal N}^{-1} \begin{pmatrix}
d^\n\\
b
\end{pmatrix},
\end{align}
and by inverting ${\cal N}^{-1}$ we find the following non-vanishing propagators
\begin{align}\label{eq:prop_ghost1}
&\langle  \bar{c}_\n(-q)d^\m(q)\rangle=-2G\left(\frac{(1+z)q^\m q_\n}{z} -\delta^\m_\n\right)\frac{i}{q^2}, \\
\label{eq:prop_ghost2}&\langle  \bar{b}(-q)d^\m(q)\rangle=-\frac{2G q^\m}{z q^4}, \\
\label{eq:prop_ghost3}&\langle \bar{b}(-q) b(q)\rangle = -\frac{G}{1+4\tau}\frac{i}{q^2}.
\end{align}
Although we could use $z$ to try to simplify the form of the propagator 
$\langle d^\m(-q) \bar{c}_\n(q)\rangle$ we prefer to keep it arbitrary in order to track gauge independence of our results.
\section{Computation of Correlation Functions}\label{sec:computation}
Once we have set-up the perturbative expansion of the action and derived the propagators, we can affront the computation of the one-loop RG flow of the different coupling constants in the Lagrangian. In order to understand what we need to compute, let us take a look to the zeroth order action around the background metric
\begin{align}
\nonumber S=\int d^4x\bigg\{ &|\bar g|^{\frac{1}{4}}\left[ -\frac{1}{2G}  \left(\bar R + \frac{3}{32}\frac{\bar \nabla_\m |\bar g| \bar \nabla^\m |\bar g|}{|\bar g|^2}\right) + \frac{1}{2}\partial_\m \phi \partial^\m \phi -\frac{\xi}{2}\phi^2 \left(\bar R + \frac{3 \bar 
\square |\bar{g}|}{4|\bar g|} - \frac{27 \bar \nabla_{\m} |\bar g| \bar \nabla^{\m}|\bar g|}{32 |\bar g|^2}\right)\right]-\frac{m^2}{2}\phi^2 -\lambda \phi^4\bigg\}.
\end{align}

By expanding this around flat space
\begin{align}
\bar{g}_{\m\n}=\eta_{\m\n}+H_{\m\n},
\end{align}
we see that it is enough to compute the two-point function of $H_{\m\n}$ in order to capture the running of $G$, while from the two and four-point functions of the scalar field we derive the running of $m^2$, $\lambda$ and the field strength renormalization of $\phi$, as usual. Finally, from the coupling $\phi^2 H_{\m\n}$ we can extract the running of $\xi$. Of course, since the theory is non-renormalizable, we will also find extra divergences corresponding to higher dimension operators -- with four derivatives. Therefore, we will adopt an EFT approach to quantization from now on.

We will perform the computation with standard Feynman diagrams, using the propagators (\ref{eq:proph}-\ref{eq:prop_phi}) and (\ref{eq:prop_ghost1}-\ref{eq:prop_ghost3}). The interaction vertices are defined in the standard way by variational derivatives of the action $S_{J}$, after expanding the background metric around flat space and going to momentum space
\begin{align}
\nonumber &\langle H_{\m_1\n_1}(q_1)\dots H_{\m_n\n_n}(q_n)h_{\a_1\b_1}(p_1)\dots h_{\a_m\b_m}(p_m)\phi(k_1)\phi(k_s)\rangle\\
&=\frac{i}{n!m!s!}\frac{\delta}{\delta H_{\m_1\n_1}(q_1)}\dots\frac{\delta}{\delta H_{\m_n\n_n}(q_n)}\frac{\delta}{\delta h_{\a_1\b_1}(p_1)}\dots\frac{\delta}{\delta h_{\a_m\b_m}(p_m)}\frac{\delta}{\delta \phi(k_1)}\dots\frac{\delta S_J}{\delta \phi(k_s)}.
\end{align}

The explicit formulas for all the vertices are pretty cumbersome and not illuminating at all, so we refrain to show them here explicitly. Let us note however that, due to background Weyl invariance, all vertices and all correlation functions that we will compute must satisfy the Ward identities \eqref{eq:Ward_identity}.

Regarding loop integrals, we have two possible poles that can enter into the loops from the propagators (\ref{eq:proph}-\ref{eq:prop_phi}) and (\ref{eq:prop_ghost1}-\ref{eq:prop_ghost3}). They represent the massless pole of the graviton and ghosts and the massive pole of the scalar field
\begin{align}
{\cal P}_{0}(q)=\frac{1}{q^2},\quad {\cal P}_{m}(q)=\frac{1}{q^2-m^2}.
\end{align}

This implies that the denominator in a typical Feynman diagram will be a product of these poles evaluated for the momentum structures running in the loops, that will depend on the external momentum $p^\m$. For example, a fish diagram will have a typical form
\begin{align}\label{eq:prototypical_diagram}
\begin{fmffile}{fish} 
\parbox{15mm}{
\begin{fmfgraph*}(80,80) 
\fmfleft{l1} 
\fmfright{r1} 
\fmf{plain}{l1,c}  
\fmf{plain}{d,r1}  
\fmf{plain,left,tension=.4}{c,d}  
\fmf{plain,left,tension=.4}{d,c}  
\end{fmfgraph*}
}\end{fmffile} \qquad \qquad \sim \int \frac{d^4k}{(2\pi)^4}\  F(p,k)\ {\cal P}_i (k+p) {\cal P}_j(k),
\end{align}
where the form-factor $F(p,k)$ will depend on the particular diagram, and we would have to choose later the pole structures depending if the internal legs are scalars or gravitons.

In the following we will be interested in the computation of divergences, which are the only piece needed to obtain the RG flow of the coupling constants. Therefore we will ignore the finite parts of the diagrams and will capture these divergences by expanding the integrands of the different diagrams in powers of the external momentum and the mass $m$ of the scalar field. After reducing any index structure as usual by using rotational invariance\footnote{When expanding the denominators in the Feynman integrals, we will encounter an increasing number of loop momenta $q^\mu$ in the numerators. We will reduce those by Lorentz (rotational) invariance in the standard way, averaging over directions \cite{Collins:1984xc},
\begin{align*}
q_{i_1}q_{i_2}\dots q_{i_n}\rightarrow |q|^n T_{i_1i_2 \dots i_n} \frac{\Gamma\left(\frac{d}{2}\right)\Gamma\left(\frac{n+1}{2}\right)}{\Gamma\left(\frac{1}{2}\right)\Gamma \left(\frac{d+n}{2}\right)}
\end{align*}
where $d$ is the space-time dimension and
\begin{align*}
T_{i_1 i_2\dots i_n}=\frac{1}{n!}\left[ \delta_{i_1i_2}\dots \delta_{i_{n-1}i_n}+\text{all permutations of the i's}\right]
\end{align*}
for even $n$, and $T_{i_1 i_2...i_n}=0$ for odd $n$.

Note that the maximum number of free loop momenta that we can find is tied to the number of indices in the external legs of the diagram. Two for every $H_{\mu\nu}$ in a external leg and one for every $p^\mu$. This means that, for example, for the two-point function of the scalar field it is enough to retain terms with up to four free $q^\mu$ (since we can have divergences proportional to $p^4$), while this amount is doubled for the graviton two-point function (four momenta and four indices in the gravitons).}, all divergent integrals in the expansion will have the same form
\begin{align}
D(n)=\int \frac{d^4 q}{(2\pi)^4}\frac{1}{q^n}.
\end{align}

Once they are taken to this form, we will use dimensional regularization in order to compute them. Since the above integrals have no dimensionful parameter, we can directly see that all of them must vanish -- as it is usual in dimensional regularization -- unless $n=4$, so we will only need to retain these integrals
\begin{align}\label{eq:integral}
{\cal I}=D(4)=\int \frac{d^4 q}{(2\pi)^4}\frac{1}{q^4}.
\end{align}

Although we are only interested in the UV divergences of the integral, we must note that \eqref{eq:integral} is however divergent on both ends of the integral. Therefore, it will be convenient for us to regulate the intermediate IR divergences by introducing a soft mass $\eta$ and rewriting ${\cal I}$ in $d$ dimensions as
\begin{align}
{\cal I}=\int \frac{d^dq}{(2\pi)^4}\frac{1}{(q^2-\eta^2)^2},
\end{align}
which can now be computed by using standard formulas to give
\begin{align}\label{eq:I_integral}
{\cal I}=\frac{i}{8\pi^2 \epsilon}-\frac{i}{16\pi^2}\left(\gamma-\log(4\pi)+\log(\eta^2)\right)+{\cal O}(\epsilon),
\end{align}
where $\gamma$ is the Euler-Mascheroni constant and $\epsilon=4-d$. This will be the form that we will later use to regularize the divergences in the Feynman diagrams. From now on we will only focus on those diagrams with non-vanishing divergences under this regularization scheme.

All the computations presented here have been performed with two independent computer codes based on Mathematica, with the help of the package xAct \cite{Brizuela:2008ra,Nutma:2013zea}; and FORM \cite{Vermaseren:2000nd}.

\subsection{The Two-point Function of the Scalar Field}
We start by computing the simplest of the correlation functions that we will need to define the RG flow of the coupling constants. That is the two-point function of the scalar field, which will be given by the following Feynman diagrams
\begin{align}
\nonumber \\
\langle \phi(-p)\phi(p)\rangle_{\rm 1-loop} = 
\begin{fmffile}{S201} 
\parbox{15mm}{
\begin{fmfgraph*}(50,50) 
\fmfleft{l1} 
\fmfright{r1} 
\fmf{plain}{l1,c}  
\fmf{plain}{c,r1}  
\fmf{plain,tension=0.4}{c,c}  
\end{fmfgraph*}
}\end{fmffile}\quad+
\begin{fmffile}{S212} 
\parbox{15mm}{
\begin{fmfgraph*}(80,80) 
\fmfleft{l1} 
\fmfright{r1} 
\fmf{plain}{l1,c}  
\fmf{plain}{d,r1}  
\fmf{plain,left,tension=.4}{c,d}  
\fmf{photon,left,tension=.4}{d,c}  
\end{fmfgraph*}
}\end{fmffile}\qquad \qquad ,
\end{align}
where our dictionary for the lines of the diagrams is shown in Table \ref{tab:dictionary}.

\begin{table}[]
\begin{tabular}{ccc}
$H_{\m\n}\equiv\begin{fmffile}{H} 
\parbox{15mm}{
\begin{fmfgraph*}(40,40) 
\fmfleft{l1} 
\fmfright{r1} 
\fmf{gluon}{l1,r1}
\end{fmfgraph*}
}\end{fmffile}\qquad \qquad$ &$h_{\m\n}\equiv\begin{fmffile}{hs} 
\parbox{15mm}{
\begin{fmfgraph*}(40,40) 
\fmfleft{l1} 
\fmfright{r1} 
\fmf{photon}{l1,r1}
\end{fmfgraph*}
}\end{fmffile}\qquad\qquad $  &$\phi\equiv\begin{fmffile}{phi} 
\parbox{15mm}{
\begin{fmfgraph*}(40,40) 
\fmfleft{l1} 
\fmfright{r1} 
\fmf{plain}{l1,r1}
\end{fmfgraph*}
}\end{fmffile}\qquad\qquad $ \\
$l\equiv\begin{fmffile}{l} 
\parbox{15mm}{
\begin{fmfgraph*}(40,40) 
\fmfleft{l1} 
\fmfright{r1} 
\fmf{dashes}{l1,r1}
\end{fmfgraph*}
}\end{fmffile}\qquad \qquad$ &$\bar{x}\equiv\begin{fmffile}{barx} 
\parbox{15mm}{
\begin{fmfgraph*}(40,40) 
\fmfleft{l1} 
\fmfright{r1} 
\fmf{dots}{l1,r1}
\end{fmfgraph*}
}\end{fmffile}\qquad\qquad $  &$x\equiv\begin{fmffile}{x} 
\parbox{15mm}{
\begin{fmfgraph*}(40,40) 
\fmfleft{l1} 
\fmfright{r1} 
\fmf{dbl_dots}{l1,r1}
\end{fmfgraph*}
}\end{fmffile}\qquad\qquad $ \\
$\alpha\equiv\begin{fmffile}{alpha} 
\parbox{15mm}{
\begin{fmfgraph*}(40,40) 
\fmfleft{l1} 
\fmfright{r1} 
\fmf{dbl_dashes}{l1,r1}
\end{fmfgraph*}
}\end{fmffile}\qquad \qquad$ &$d^\m, \bar{c}_\m \equiv\begin{fmffile}{ghosts} 
\parbox{15mm}{
\begin{fmfgraph*}(40,40) 
\fmfleft{l1} 
\fmfright{r1} 
\fmf{fermion}{l1,r1}
\end{fmfgraph*}
}\end{fmffile}\qquad\qquad $  &$b,\bar{b}\equiv\begin{fmffile}{ghostsb} 
\parbox{15mm}{
\begin{fmfgraph*}(40,40) 
\fmfleft{l1} 
\fmfright{r1} 
\fmf{dots_arrow}{l1,r1}
\end{fmfgraph*}
}\end{fmffile}\qquad\qquad $ 
\end{tabular}
\caption{Dictionary of lines for the Feynman diagrams.} \label{tab:dictionary}
\end{table}

By inspection of the action and the topology of the diagrams, we see that we can expect three types of divergences, proportional to $p^0$, $p^2$ and $p^4$. In principle, we could have added a gravitational tadpole here. However, it is proportional to the integral $D(2)$ and therefore it vanishes in dimensional regularization. 

Computing the diagrams as previously discussed, expanding the denominators, and retaining only the UV divergent terms we find
\begin{align}\label{eq:2pt_scalar}
\langle \phi(-p)\phi(p)\rangle_{\rm 1-loop}= \left(\frac{G p^4 (3+2\sigma (3+8\xi))}{4\sigma}-\frac{3 G m^2 p^2 (1+2\sigma (1+8\xi +4\xi^2)}{4\sigma}+12\lambda m^2 - 6 G m^4 \xi^2\right){\cal I},
\end{align}
which indeed contains the three possible divergences previously mentioned. From the momentum-independent term we will be able to extract the running of $m^2$, while the term proportional to $p^2$ will give the field strength renormalization of the field. The piece quartic in the external momentum will require the introduction of a higher-derivative operator in order to absorb the divergence, as usual in a non-renormalizable EFT.

\subsection{The Four-point Function of the Scalar Field}
We compute now the four-point function of the scalar field. As before, we expect divergences with external momentum up to $p^4$. The corresponding Feynman diagrams contributing to this are
\begin{align}
\langle \phi(-p)\phi(-p)\phi(p)\phi(p)  \rangle_{\rm 1-loop}=  
\begin{fmffile}{4p0} 
\parbox{15mm}{
\begin{fmfgraph*}(50,50) 
\fmfleft{l1,l2} 
\fmfright{r1,r2} 
\fmf{plain}{l1,c}  
\fmf{plain}{l2,c}  
\fmf{plain}{d,r1}  
\fmf{plain}{d,r2} 
\fmf{plain,left,tension=0.4}{c,d,c}  
\end{fmfgraph*}
}\end{fmffile}\quad  +
\begin{fmffile}{4p1} 
\parbox{15mm}{
\begin{fmfgraph*}(50,50) 
\fmfleft{l1,l2} 
\fmfright{r1,r2} 
\fmf{plain}{l1,c}  
\fmf{plain}{l2,c}  
\fmf{plain}{d,r1}  
\fmf{plain}{d,r2} 
\fmf{photon,left,tension=0.4}{c,d,c}  
\end{fmfgraph*}
}\end{fmffile}\quad +
\begin{fmffile}{4p2} 
\parbox{15mm}{
\begin{fmfgraph*}(50,50) 
\fmfbottom{l1,l2} 
\fmftop{r1,r2} 
\fmf{plain}{l1,c1}  
\fmf{plain}{l2,c2}  
\fmf{plain}{r1,c3}  
\fmf{plain}{r2,c3}
\fmf{photon,tension=.2}{c1,c3,c2}
\fmf{plain,tension=.2}{c1,c2}  
\end{fmfgraph*}
}\end{fmffile}\quad+
\begin{fmffile}{4p3} 
\parbox{15mm}{
\begin{fmfgraph*}(50,50) 
\fmfbottom{l1,l2} 
\fmftop{r1,r2} 
\fmf{plain}{l1,c1}  
\fmf{plain}{l2,c2}  
\fmf{plain}{r1,c3}  
\fmf{plain}{r2,c3}
\fmf{plain,tension=.2}{c1,c3,c2}
\fmf{photon,tension=.2}{c1,c2}  
\end{fmfgraph*}
}\end{fmffile}\quad +
\begin{fmffile}{4p4} 
\parbox{15mm}{
\begin{fmfgraph*}(50,50) 
\fmfleft{l1,l2} 
\fmfright{r1,r2} 
\fmf{plain}{l1,c1}  
\fmf{plain}{l2,c2}  
\fmf{plain}{r1,c3}  
\fmf{plain}{r2,c4}
\fmf{photon,tension=.5}{c1,c3}
\fmf{photon,tension=.5}{c2,c4}
\fmf{plain,tension=.5}{c1,c2}  
\fmf{plain,tension=.5}{c3,c4}  
\end{fmfgraph*}
}\end{fmffile}\qquad ,
\end{align}
where we are just drawing inequivalent topologies. For all the diagrams considered here, we must sum the contribution of all inequivalent channels once the external momenta are fixed. This amounts to adding the $s$, $t$ and $u$ channels for all the diagrams, plus two permutations of the external vertices for the triangles, which add up to six different channels. 

We evaluate the divergences by setting the magnitude of all external momenta to that of $p^\m$. This is is equivalent to the kinematical configuration $s=4p^2$, $t=u=0$, which will define our subtraction point. Under this choice, the one-loop contribution to the four-point function becomes
\begin{align}\label{eq:4-pt_scalar}
\nonumber &\langle \phi(-p)\phi(-p)\phi(p)\phi(p)  \rangle_{\rm 1-loop}=\left[36 (24 \lambda^2  -48 G \lambda m^2 \xi ^2 + G^2 m^4 \xi ^3 (2 + 9 \xi ))\right. \\
\nonumber &+\frac{6 G p^2 (-6 \lambda (1 + \sigma (2 + 8 \xi  (4 + \xi ))) + G 
m^2 \xi  (-3 \xi  + \sigma (1 + 2 \xi  (11 + 6 \xi  (7 + 3 \xi \
)))))}{\sigma}\\
&\left. +\frac{G^2 p^4 (117 + 8 \xi  (17 + 40 \xi )  -4 \sigma (27 + 4 \xi  
(37 + 45 \xi )) + 4 \sigma^2 (483 + 4 \xi  (259 + 2 \xi  (385 + 3 \xi  
(-28 + 9 \xi )))))}{8 \sigma^2}\right]{\cal I}.
\end{align}

As in the previous section, we obtain three kind of divergences. The momentum independent one will dictate the running of $\lambda$, while the other terms will demand higher-derivative operators to be introduced in the EFT expansion.

\subsection{Corrections to the Non-minimal Coupling}
In order to compute the one-loop contribution to the running of the non-minimal coupling we will need to focus on the three-point function mixing two external scalar fields and a graviton. The tree-level form of this correlator can be obtained by expanding the action to the given order in the background graviton, giving
\begin{align}\label{eq:non-minimal_tree}
\langle \phi(-p)\phi(-p)H_{\m\n}(2p)\rangle_{\rm tree}=- \frac{i}{4} (1 + 4\xi ) ( p^2 \eta_{\m\n} -4 p_{\m} p_{\n}),
\end{align}
where we have assigned equal incoming momentum for the scalar fields.

Therefore, contributions to $\langle \phi(-p)\phi(-p)H_{\m\n}(2p)\rangle$ will renormalize the combination $1+4\xi$, once the effect of the field strength renormalization of $\phi$  is subtracted. Note that, since the theory is Weyl invariant at the background level, the action must satisfy the condition \eqref{eq:Ward_identity}, which implies that $\langle \phi(-p)\phi(-p)H_{\m\n}(p)\rangle$ must be a traceless tensor. This is trivially satisfied by the tree-level contribution \eqref{eq:non-minimal_tree} but it will serve as a strong sanity check of our result for the one-loop computation since in that case the condition is satisfied in a non-trivial way.

The one-loop topologies contributing to this correlator are
\begin{align}
\nonumber \langle \phi(-p)\phi(-p)H_{\m\n}(2p)\rangle_{\rm 1-loop} &=
\begin{fmffile}{nm1} 
\parbox{15mm}{
\begin{fmfgraph*}(50,50) 
\fmfbottom{l1,l2} 
\fmftop{r1} 
\fmf{plain}{l1,c1}  
\fmf{plain}{l2,c2}  
\fmf{gluon}{r1,c3}  
\fmf{photon,tension=.2}{c1,c3,c2}
\fmf{plain,tension=.2}{c1,c2}  
\end{fmfgraph*}
}\end{fmffile}\quad+
\begin{fmffile}{nm2} 
\parbox{15mm}{
\begin{fmfgraph*}(50,50) 
\fmfbottom{l1,l2} 
\fmftop{r1} 
\fmf{plain}{l1,c1}  
\fmf{plain}{l2,c2}  
\fmf{gluon}{r1,c3}  
\fmf{plain,tension=.2}{c1,c3,c2}
\fmf{photon,tension=.2}{c1,c2}  
\end{fmfgraph*}
}\end{fmffile}\quad+
\begin{fmffile}{nm3} 
\parbox{15mm}{
\begin{fmfgraph*}(50,50) 
\fmftop{t}
\fmfbottom{b}
\fmfleft{l,l2}
\fmfright{r}
\fmf{plain}{l,li}
\fmf{plain}{l2,li}
\fmf{gluon}{ri,r}
\fmf{phantom}{t,ti}
\fmf{phantom}{bi,b}
\fmf{photon,left=.4,tension=0.4}{ri,bi,li,ti,ri}  
\end{fmfgraph*}
}\end{fmffile}\qquad  +
\begin{fmffile}{nm4} 
\parbox{15mm}{
\begin{fmfgraph*}(50,50) 
\fmftop{t}
\fmfbottom{b}
\fmfleft{l,l2}
\fmfright{r}
\fmf{plain}{l,li}
\fmf{plain}{l2,li}
\fmf{gluon}{ri,r}
\fmf{phantom}{t,ti}
\fmf{phantom}{bi,b}
\fmf{plain,left=.4,tension=0.4}{ri,bi,li,ti,ri}  
\end{fmfgraph*}
}\end{fmffile}\\
&+\begin{fmffile}{nm5} 
\parbox{15mm}{
\begin{fmfgraph*}(50,50) 
\fmftop{t}
\fmfbottom{b}
\fmfleft{l,l2}
\fmfright{r}
\fmf{plain}{l,li}
\fmf{gluon}{l2,li}
\fmf{plain}{ri,r}
\fmf{phantom}{t,ti}
\fmf{phantom}{bi,b}
\fmf{photon,left=.4,tension=0.4}{ri,bi,li}  
\fmf{plain,left=.4,tension=0.4}{li,ti,ri}  
\end{fmfgraph*}
}\end{fmffile} \qquad  +
\begin{fmffile}{nm6} 
\parbox{15mm}{
\begin{fmfgraph*}(50,50) 
\fmftop{t}
\fmfbottom{b}
\fmfleft{l,l2}
\fmfright{r}
\fmfv{decor.shape=circle,decor.filled=shaded,decor.size=6}{ti}
\fmf{plain}{l,li}
\fmf{plain}{l2,li}
\fmf{gluon}{ri,r}
\fmf{phantom}{t,ti}
\fmf{phantom}{bi,b}
\fmf{photon,left=.4,tension=0.4}{ri,bi,li,ti}  
\fmf{dashes,left=.4,tension=0.4}{ti,ri}  
\end{fmfgraph*}
}\end{fmffile}  \qquad  +
\begin{fmffile}{nm7} 
\parbox{15mm}{
\begin{fmfgraph*}(50,50) 
\fmftop{t}
\fmfbottom{b}
\fmfleft{l,l2}
\fmfright{r}
\fmfv{decor.shape=circle,decor.filled=shaded,decor.size=6}{ti}
\fmf{plain}{l,li}
\fmf{plain}{l2,li}
\fmf{gluon}{ri,r}
\fmf{phantom}{t,ti}
\fmf{phantom}{bi,b}
\fmf{photon,left=.4,tension=0.4}{ri,bi,li,ti}  
\fmf{dots,left=.4,tension=0.4}{ti,ri}  
\end{fmfgraph*}
}\end{fmffile}\quad ,
\end{align}
where the last two diagrams contain a explicit presence of the bosonic ghost fields in the internal lines, with the small shaded blown representing kinetic mixing. Actually, the presence of these bosonic ghost fields is critical, since the sum of all the other diagrams is \emph{not} traceless and therefore violates the Ward identity \eqref{eq:Ward_identity}. It is only when the last two topologies, which also have a non-vanishing trace, are added, that the whole contribution becomes traceless. This is not surprising, of course. The role of ghosts is precisely to cancel the dynamics of gauge modes, which violate Ward identities, in the internal legs. However, it serves here as a very non-trivial test of the construction of the path integral, the BRST sector and of our computation.

The final result takes the form
\begin{align}\label{eq:non-minimal_one-loop}
\nonumber \langle \phi(-p)\phi(-p)H_{\m\n}(2p)\rangle_{\rm 1-loop} ={\cal I}\left(p_\m p_\n-\frac{1}{4}p^2 \eta_{\m\n}\right)&\left[- \frac{64 \lambda \sigma (1 + 6 \xi )  - G m^2 (-3 + 2 \sigma (1 + 6 \xi ) (-3 + 2 \xi  (-3 + 8 \xi )))}{4 \sigma}\right.\\
&\left. +\frac{G p^2 (-9  -40 \xi  + 2 \sigma (23 + 30 \xi  + \sigma (42 + 4 \xi  (53 + 6 (13  -6 \xi ) \xi ))))}{12 \sigma^2}\right].
\end{align}

We see that the result is indeed proportional to \eqref{eq:non-minimal_tree}. Moreover, no terms independent of the momentum have been generated. Those would require the introduction of counter-terms of the schematic form $|g|^{\alpha} \phi^2$, with $\alpha$ a constant, that violate Weyl invariance.


\subsection{The Gravitational Two-point Function}
The last correlation function that we need in order to compute the RG flow of the coupling constants in the action is the two-point function of the gravitational field. Its value is required in order to get the running of the Newton constant G. Additionally, we will also compute the contributions that require the introduction of higher-derivative operators to cancel divergences. This will not only complete our computation but it will also serve as a third additional computation complementary to that of \cite{Alvarez:2015sba,Ardon:2017atk}. In the following we will split the computation in three parts -- the contribution of the scalar field, that of the rest of bosonic fields, and the one coming from ghost loops.

\subsubsection{Contributions from scalar loops}
This first contribution in the simplest one of all that we will consider in this subsection. It is equivalent to compute the contribution of a gravitating scalar-field in a background non-dynamical geometry. As such, and by the reasons discussed in this work, its contribution shall be identical to that coming from GR. Indeed, we have checked that it is the case at the level of $\beta$-functions.

There are only two diagrams that need to be taken into account
\begin{align}
\nonumber \\
\langle H_{\m\n}(-p) H_{\a\b}(p)\rangle_{\phi}=\quad 
\begin{fmffile}{G201} 
\parbox{15mm}{
\begin{fmfgraph*}(50,50) 
\fmfleft{l1} 
\fmfright{r1} 
\fmf{gluon}{l1,c}  
\fmf{gluon}{c,r1}  
\fmf{plain,tension=0.4}{c,c}  
\end{fmfgraph*}
}\end{fmffile}\qquad +\quad  
\begin{fmffile}{G202} 
\parbox{15mm}{
\begin{fmfgraph*}(50,50) 
\fmftop{t}
\fmfbottom{b}
\fmfleft{l}
\fmfright{r}
\fmf{gluon}{l,li}
\fmf{gluon}{ri,r}
\fmf{phantom}{t,ti}
\fmf{phantom}{bi,b}
\fmf{plain,left=.4,tension=0.4}{ri,bi,li,ti,ri}   
\end{fmfgraph*}
}\end{fmffile}\quad ,
\end{align}
and whose contribution is
\begin{align}
\nonumber \langle H_{\m\n}(-p) H_{\a\b}(p)\rangle_{\phi}&=  \left[\frac{p^4}{480}  (-1 + 20 \xi  + 60 \xi ^2) \eta_{\a\b} 
\eta_{\m\n}  - \frac{p^2}{120}  (1 + 20 \xi  + 60 \xi ^2)\left( \eta_{\m\n}
p_{\a} p_{\b} + \eta_{\a\b} p_{\m} p_{\n}\right) \right. \\
\nonumber & - \frac{p^2}{120}\left( \eta_{\b\n} p_{\a} p_{\m} + \eta_{\a\n} p_{\b} p_{\m}  +\eta_{\b\m} p_{\a} p_{\n} + \eta_{\a\m} p_{\b} p_{\n}\right)  + \left(\frac{1}{15} + \frac{2\xi}{3}  + 2 \xi ^2\right) p_{\a} p_{\b} p_{\m} p_{\n}\\
\nonumber &+\frac{p^4}{120}\left( \eta_{\a\n} \eta_{\b\m} + \eta_{\a\m} 
\eta_{\b\n}\right)-\frac{m^2(1+6\xi)}{48}\left(4 p^2 (\eta_{\a\n} \eta_{\b\m} +\eta_{\a\m} \eta_{\b\n})  + 4 (\eta_{\m\n} p_{\a} p_{\b} +\eta_{\a\b} p_{\m} p_{\n})\right.\\
&\left. \left. -4( \eta_{\b\n} p_{\a} p_{\m}+ \eta_{\a\n}p_{\b} p_{\m} + \eta_{\b\m} p_{\a} p_{\n} +\eta_{\a\m} p_{\b} p_{\n})-3 p^2 \eta_{\a\b} \eta_{\m\n}\right)\right]{\cal I}.
\end{align}

As in the case of the non-minimal coupling, note that there are no terms independent of the external momentum, since those would imply a renormalization of the cosmological constant, violating Weyl invariance of the background. The satisfaction of the Ward identity \eqref{eq:Ward_identity} can be seen here from the fact that
\begin{align}
\langle H_{\m\n}(-p) H_{\a\b}(p)\rangle_{\phi}\eta^{\m\n}\eta^{\a\b}=0.
\end{align}

The terms proportional to $p^2$ will renormalize the Newton's constant $G$ -- as it can be seen from the fact that they are proportional to the tree-level kinetic term of $H_{\m\n}$ --, while the terms with a quartic dependence on $p^4$ will require higher-derivative terms.

\subsubsection{Contributions from the graviton and bosonic ghost fields}
While the contribution from the scalar field to the gravitational two-point function is pretty simple, that of the rest of bosonic fields is pretty cumbersome, due to the kinetic mixing between the graviton fluctuation $h_{\m\n}$ and the bosonic ghosts $l$ and $\bar{x}$. This multiplies the number of Feynman diagrams to be considered and leaves the following set of inequivalent topologies
\begin{align}
\langle H_{\m\n}&(-p)H_{\a\b}(p)\rangle_{\rm bosons}=
\nonumber  \ \ \ \begin{fmffile}{b1} 
\parbox{15mm}{
\begin{fmfgraph*}(80,80) 
\fmftop{t}
\fmfbottom{b}
\fmfleft{l}
\fmfright{r}
\fmf{gluon}{l,li}
\fmf{gluon}{ri,r}
\fmf{phantom}{t,ti}
\fmf{phantom}{bi,b}
\fmf{photon,left=.4,tension=0.4}{li,ti,ri,bi,li}  
\end{fmfgraph*}
}\end{fmffile}\qquad  \qquad  +
\begin{fmffile}{b2a} 
\parbox{15mm}{
\begin{fmfgraph*}(80,80) 
\fmftop{t}
\fmfbottom{b}
\fmfleft{l}
\fmfright{r}
\fmfv{decor.shape=circle,decor.filled=shaded,decor.size=6}{ti}
\fmf{gluon}{l,li}
\fmf{gluon}{ri,r}
\fmf{phantom}{t,ti}
\fmf{phantom}{bi,b}
\fmf{photon,left=.4,tension=0.4}{ri,bi,li,ti}  
\fmf{dashes,left=.4,tension=0.4}{ti,ri}  
\end{fmfgraph*}
}\end{fmffile} \qquad  \qquad  +
 \begin{fmffile}{b2b} 
\parbox{15mm}{
\begin{fmfgraph*}(80,80) 
\fmftop{t}
\fmfbottom{b}
\fmfleft{l}
\fmfright{r}
\fmfv{decor.shape=circle,decor.filled=shaded,decor.size=6}{ti}
\fmf{gluon}{l,li}
\fmf{gluon}{ri,r}
\fmf{phantom}{t,ti}
\fmf{phantom}{bi,b}
\fmf{photon,left=.4,tension=0.4}{ri,bi,li,ti}  
\fmf{dots,left=.4,tension=0.4}{ti,ri}  
\end{fmfgraph*}
}\end{fmffile}\\
&+ \begin{fmffile}{b3a} 
\parbox{15mm}{
\begin{fmfgraph*}(80,80) 
\fmftop{t}
\fmfbottom{b}
\fmfleft{l}
\fmfright{r}
\fmf{gluon}{l,li}
\fmf{gluon}{ri,r}
\fmf{phantom}{t,ti}
\fmf{phantom}{bi,b}
\fmf{photon,left=.4,tension=0.4}{li,ti,ri}  
\fmf{dashes,left=.4,tension=0.4}{ri,bi,li}  
\end{fmfgraph*}
}\end{fmffile}\qquad  \qquad  +
\nonumber \begin{fmffile}{b3b} 
\parbox{15mm}{
\begin{fmfgraph*}(80,80) 
\fmftop{t}
\fmfbottom{b}
\fmfleft{l}
\fmfright{r}
\fmf{gluon}{l,li}
\fmf{gluon}{ri,r}
\fmf{phantom}{t,ti}
\fmf{phantom}{bi,b}
\fmf{photon,left=.4,tension=0.4}{li,ti,ri}  
\fmf{dots,left=.4,tension=0.4}{ri,bi,li}  
\end{fmfgraph*}
}\end{fmffile}\qquad  \qquad  +
 \begin{fmffile}{b3c} 
\parbox{15mm}{
\begin{fmfgraph*}(80,80) 
\fmftop{t}
\fmfbottom{b}
\fmfleft{l}
\fmfright{r}
\fmfv{decor.shape=circle,decor.filled=shaded,decor.size=6}{bi}
\fmf{gluon}{l,li}
\fmf{gluon}{ri,r}
\fmf{phantom}{t,ti}
\fmf{phantom}{bi,b}
\fmf{photon,left=.4,tension=0.4}{li,ti,ri}  
\fmf{dashes,right=.4,tension=0.4}{li,bi} 
\fmf{dots,right=.4,tension=0.4}{bi,ri}  
\end{fmfgraph*}
}\end{fmffile}\qquad  \qquad  +
 \begin{fmffile}{b3d} 
\parbox{15mm}{
\begin{fmfgraph*}(80,80) 
\fmftop{t}
\fmfbottom{b}
\fmfleft{l}
\fmfright{r}
\fmfv{decor.shape=circle,decor.filled=shaded,decor.size=6}{bi}
\fmfv{decor.shape=circle,decor.filled=shaded,decor.size=6}{ti}
\fmf{gluon}{l,li}
\fmf{gluon}{ri,r}
\fmf{phantom}{t,ti}
\fmf{phantom}{bi,b}
\fmf{photon,left=.4,tension=0.4}{li,ti}  
\fmf{photon,right=.4,tension=0.4}{li,bi} 
\fmf{dashes,right=.4,tension=0.4}{bi,ri}  
\fmf{dashes,left=.4,tension=0.4}{ti,ri}
\end{fmfgraph*}
}\end{fmffile}\\
 &+ \begin{fmffile}{b3e} 
\parbox{15mm}{
\begin{fmfgraph*}(80,80) 
\fmftop{t}
\fmfbottom{b}
\fmfleft{l}
\fmfright{r}
\fmfv{decor.shape=circle,decor.filled=shaded,decor.size=6}{bi}
\fmfv{decor.shape=circle,decor.filled=shaded,decor.size=6}{ti}
\fmf{gluon}{l,li}
\fmf{gluon}{ri,r}
\fmf{phantom}{t,ti}
\fmf{phantom}{bi,b}
\fmf{photon,left=.4,tension=0.4}{li,ti}  
\fmf{photon,right=.4,tension=0.4}{li,bi} 
\fmf{dots,right=.4,tension=0.4}{bi,ri}  
\fmf{dots,left=.4,tension=0.4}{ti,ri}
\end{fmfgraph*}
}\end{fmffile}\qquad \qquad +
\nonumber \begin{fmffile}{b3f} 
\parbox{15mm}{
\begin{fmfgraph*}(80,80) 
\fmftop{t}
\fmfbottom{b}
\fmfleft{l}
\fmfright{r}
\fmfv{decor.shape=circle,decor.filled=shaded,decor.size=6}{bi}
\fmfv{decor.shape=circle,decor.filled=shaded,decor.size=6}{ti}
\fmf{gluon}{l,li}
\fmf{gluon}{ri,r}
\fmf{phantom}{t,ti}
\fmf{phantom}{bi,b}
\fmf{photon,left=.4,tension=0.4}{li,ti}  
\fmf{dashes,right=.4,tension=0.4}{li,bi} 
\fmf{photon,right=.4,tension=0.4}{bi,ri}  
\fmf{dashes,left=.4,tension=0.4}{ti,ri}
\end{fmfgraph*}
}\end{fmffile}\qquad  \qquad  +
\begin{fmffile}{b3g} 
\parbox{15mm}{
\begin{fmfgraph*}(80,80) 
\fmftop{t}
\fmfbottom{b}
\fmfleft{l}
\fmfright{r}
\fmfv{decor.shape=circle,decor.filled=shaded,decor.size=6}{bi}
\fmfv{decor.shape=circle,decor.filled=shaded,decor.size=6}{ti}
\fmf{gluon}{l,li}
\fmf{gluon}{ri,r}
\fmf{phantom}{t,ti}
\fmf{phantom}{bi,b}
\fmf{photon,left=.4,tension=0.4}{li,ti}  
\fmf{dots,right=.4,tension=0.4}{li,bi} 
\fmf{photon,right=.4,tension=0.4}{bi,ri}  
\fmf{dots,left=.4,tension=0.4}{ti,ri}
\end{fmfgraph*}
}\end{fmffile}\qquad  \qquad  +
 \begin{fmffile}{b3h} 
\parbox{15mm}{
\begin{fmfgraph*}(80,80) 
\fmftop{t}
\fmfbottom{b}
\fmfleft{l}
\fmfright{r}
\fmfv{decor.shape=circle,decor.filled=shaded,decor.size=6}{bi}
\fmfv{decor.shape=circle,decor.filled=shaded,decor.size=6}{ti}
\fmf{gluon}{l,li}
\fmf{gluon}{ri,r}
\fmf{phantom}{t,ti}
\fmf{phantom}{bi,b}
\fmf{photon,left=.4,tension=0.4}{li,ti}  
\fmf{dots,right=.4,tension=0.4}{li,bi} 
\fmf{photon,right=.4,tension=0.4}{bi,ri}  
\fmf{dashes,left=.4,tension=0.4}{ti,ri}
\end{fmfgraph*}
}\end{fmffile}\\
&+ \begin{fmffile}{b4a} 
\parbox{15mm}{
\begin{fmfgraph*}(80,80) 
\fmftop{t}
\fmfbottom{b}
\fmfleft{l}
\fmfright{r}
\fmfv{decor.shape=circle,decor.filled=shaded,decor.size=6}{ti}
\fmf{gluon}{l,li}
\fmf{gluon}{ri,r}
\fmf{phantom}{t,ti}
\fmf{phantom}{bi,b}
\fmf{photon,left=.4,tension=0.4}{li,ti}  
\fmf{dashes,left=.4,tension=0.4}{ti,ri,bi,li} 
\end{fmfgraph*}
}\end{fmffile}\qquad \qquad +
\begin{fmffile}{b4b} 
\parbox{15mm}{
\begin{fmfgraph*}(80,80) 
\fmftop{t}
\fmfbottom{b}
\fmfleft{l}
\fmfright{r}
\fmfv{decor.shape=circle,decor.filled=shaded,decor.size=6}{ti}
\fmf{gluon}{l,li}
\fmf{gluon}{ri,r}
\fmf{phantom}{t,ti}
\fmf{phantom}{bi,b}
\fmf{photon,left=.4,tension=0.4}{li,ti}  
\fmf{dots,left=.4,tension=0.4}{ti,ri,bi,li} 
\end{fmfgraph*}
}\end{fmffile}\qquad  \qquad  +
 \begin{fmffile}{b4c} 
\parbox{15mm}{
\begin{fmfgraph*}(80,80) 
\fmftop{t}
\fmfbottom{b}
\fmfleft{l}
\fmfright{r}
\fmfv{decor.shape=circle,decor.filled=shaded,decor.size=6}{ti}
\fmfv{decor.shape=circle,decor.filled=shaded,decor.size=6}{bi}
\fmf{gluon}{l,li}
\fmf{gluon}{ri,r}
\fmf{phantom}{t,ti}
\fmf{phantom}{bi,b}
\fmf{photon,left=.4,tension=0.4}{li,ti}  
\fmf{dots,left=.4,tension=0.4}{ti,ri,bi} 
\fmf{dashes,left=.4,tension=0.4}{bi,li} 
\end{fmfgraph*}
}\end{fmffile}\qquad  \qquad  +
 \begin{fmffile}{b4d} 
\parbox{15mm}{
\begin{fmfgraph*}(80,80) 
\fmftop{t}
\fmfbottom{b}
\fmfleft{l}
\fmfright{r}
\fmfv{decor.shape=circle,decor.filled=shaded,decor.size=6}{ti}
\fmfv{decor.shape=circle,decor.filled=shaded,decor.size=6}{bi}
\fmf{gluon}{l,li}
\fmf{gluon}{ri,r}
\fmf{phantom}{t,ti}
\fmf{phantom}{bi,b}
\fmf{photon,left=.4,tension=0.4}{li,ti}  
\fmf{dashes,left=.4,tension=0.4}{ti,ri,bi} 
\fmf{dots,left=.4,tension=0.4}{bi,li} 
\end{fmfgraph*}
}\end{fmffile}\nonumber \\
&+  \begin{fmffile}{b5a} 
\parbox{15mm}{
\begin{fmfgraph*}(80,80) 
\fmftop{t}
\fmfbottom{b}
\fmfleft{l}
\fmfright{r}
\fmf{gluon}{l,li}
\fmf{gluon}{ri,r}
\fmf{phantom}{t,ti}
\fmf{phantom}{bi,b}
\fmf{dots,left=.4,tension=0.4}{li,ti,ri,bi,li}  
\end{fmfgraph*}
}\end{fmffile}\qquad \qquad 
+ \begin{fmffile}{b5b} 
\parbox{15mm}{
\begin{fmfgraph*}(80,80) 
\fmftop{t}
\fmfbottom{b}
\fmfleft{l}
\fmfright{r}
\fmf{gluon}{l,li}
\fmf{gluon}{ri,r}
\fmf{phantom}{t,ti}
\fmf{phantom}{bi,b}
\fmf{dashes,left=.4,tension=0.4}{li,ti,ri,bi,li}  
\end{fmfgraph*}
}\end{fmffile}\qquad  \qquad  +
 \begin{fmffile}{b5c} 
\parbox{15mm}{
\begin{fmfgraph*}(80,80) 
\fmftop{t}
\fmfbottom{b}
\fmfleft{l}
\fmfright{r}
\fmfv{decor.shape=circle,decor.filled=shaded,decor.size=6}{ti}
\fmfv{decor.shape=circle,decor.filled=shaded,decor.size=6}{bi}
\fmf{gluon}{l,li}
\fmf{gluon}{ri,r}
\fmf{phantom}{t,ti}
\fmf{phantom}{bi,b}
\fmf{dashes,left=.4,tension=0.4}{bi,li,ti}  
\fmf{dots,left=.4,tension=0.4}{ti,ri,bi}  
\end{fmfgraph*}
}\end{fmffile}\qquad  \qquad  +
 \begin{fmffile}{b6} 
\parbox{15mm}{
\begin{fmfgraph*}(80,80) 
\fmftop{t}
\fmfbottom{b}
\fmfleft{l}
\fmfright{r}
\fmfv{decor.shape=circle,decor.filled=shaded,decor.size=6}{ti}
\fmfv{decor.shape=circle,decor.filled=shaded,decor.size=6}{bi}
\fmf{gluon}{l,li}
\fmf{gluon}{ri,r}
\fmf{phantom}{t,ti}
\fmf{phantom}{bi,b}
\fmf{dbl_dashes,left=.4,tension=0.4}{li,ti}  
\fmf{dbl_dots,left=.4,tension=0.4}{ti,ri}  
\fmf{dbl_dashes,left=.4,tension=0.4}{ri,bi}  
\fmf{dbl_dots,left=.4,tension=0.4}{bi,li}  
\end{fmfgraph*}
}\end{fmffile}\qquad \qquad .
\end{align}

Computing these Feynman diagrams by following the methods previously described in this paper, we find the following result
\begin{align}
\nonumber &\langle H_{\m\n}(-p)H_{\a\b}(p)\rangle_{\rm bosons}= \left(\frac{p^4 (50 + 100 \sigma + 777 \sigma^2) (\eta_{\a\m}\eta_{\b\n}+\eta_{\a\n}\eta_{\b\m})}{480 \sigma^2} - 
\frac{p^4 (125 + 250 \sigma + 2167 \sigma^2) \eta_{\a\b}\eta_{\m\n}}{1920 \sigma^2}\right. \\
\nonumber & - \frac{p^2 (50 + 100 \sigma + 777 \sigma^2)(p_\a p_\m \eta_{\b\n}+p_\a p_\n \eta_{\b\m}+p_\b p_\m \eta_{\a\n}+p_\b p_\n \eta_{\a\m})}{480 \sigma^2}+ \frac{(25 + 50 \sigma + 164 \sigma^2) p_{\a} p_{\b} p_{\m} p_{\n}}{120 \sigma^2} \\
&\left. + \frac{p^2 (25 + 50 \sigma + 613 \sigma^2) (\eta_{\m\n}p_\a p_\b +\eta_{\a\b}p_\m p_\n)}{480 \sigma^2}\right){\cal I}.
\end{align}

Note that in this case all divergences are proportional to $p^4$ as a consequence of the absence of any dimensionful parameter in the loops, since all the fields that propagate in these diagrams are massless. As a consequence, this contribution will only renormalize higher-derivative operators.

\subsubsection{Contributions from fermionic ghosts}
The last contribution that we need to compute in order to get the full one-loop divergence contributing to the gravitational two-point functions is that coming from the loops of fermionic ghosts, $\bar{c}_\m$, $d^\m$, $\bar b$ and $b$. It is given by the following diagrams
\begin{align}
\nonumber \langle H_{\m\n}(-p) H_{\a\b}(p)\rangle_{\rm fermions}=& \ \ \  \begin{fmffile}{gh1} 
\parbox{15mm}{
\begin{fmfgraph*}(80,80) 
\fmftop{t}
\fmfbottom{b}
\fmfleft{l}
\fmfright{r}
\fmf{gluon}{l,li}
\fmf{gluon}{ri,r}
\fmf{phantom}{t,ti}
\fmf{phantom}{bi,b}
\fmf{fermion,left=.4,tension=0.4}{ri,bi,li,ti,ri}  
\end{fmfgraph*}
}\end{fmffile}\qquad   \qquad +
\begin{fmffile}{gh2} 
\parbox{15mm}{
\begin{fmfgraph*}(80,80) 
\fmftop{t}
\fmfbottom{b}
\fmfleft{l}
\fmfright{r}
\fmf{gluon}{l,li}
\fmf{gluon}{ri,r}
\fmf{phantom}{t,ti}
\fmf{phantom}{bi,b}
\fmf{dots_arrow,left=.4,tension=0.4}{ri,bi,li,ti,ri}  
\end{fmfgraph*}
}\end{fmffile}\\
& +\begin{fmffile}{gh3} 
\parbox{15mm}{
\begin{fmfgraph*}(80,80) 
\fmftop{t}
\fmfbottom{b}
\fmfleft{l}
\fmfright{r}
\fmf{gluon}{l,li}
\fmf{gluon}{ri,r}
\fmf{phantom}{t,ti}
\fmf{phantom}{bi,b}
\fmf{fermion,left=.4,tension=0.4}{li,ti,ri}  
\fmf{dots_arrow,left=.4,tension=0.4}{ri,bi,li}  
\end{fmfgraph*}
}\end{fmffile}\qquad   \qquad +
\begin{fmffile}{gh4} 
\parbox{15mm}{
\begin{fmfgraph*}(80,80) 
\fmftop{t}
\fmfbottom{b}
\fmfleft{l}
\fmfright{r}
\fmfv{decor.shape=circle,decor.filled=shaded,decor.size=6}{ti}
\fmf{gluon}{l,li}
\fmf{gluon}{ri,r}
\fmf{phantom}{t,ti}
\fmf{phantom}{bi,b}
\fmf{fermion,left=.4,tension=0.4}{ri,bi,li,ti}  
\fmf{dots_arrow,left=.4,tension=0.4}{ti,ri}  
\end{fmfgraph*}
}\end{fmffile}\qquad \qquad ,
\end{align}
where the arrows indicate the fermion flow. Their contribution to the correlation function is
\begin{align}
\nonumber \langle H_{\m\n}(-p) H_{\a\b}(p)\rangle_{\rm fermions}=&\left(\frac{p^4}{16}  (\eta_{\a\m}\eta_{\b\n}+\eta_{\a\n}\eta_{\b\m})   - \frac{p^2}{16}  (p_\m p_\a \eta_{\n\b} + p_\n p_\a \eta_{\m\b} + p_\m p_\b \eta_{\n\a}+p_\n p_\b \eta_{\a\m})\right. \\
& \left.+ \frac{5}{48} p^2 (\eta_{\m\n}p_\a p_\b +\eta_{\a\b} p_\m p_\n) - \frac{1}{6}p_\a p_\b p_\m p_\n- \frac{11}{192} p^4 \eta_{\a\b}\eta_{\m\n} \right){\cal I}.
\end{align}

Again, since there are no dimensionful constant running in the loop propagators, the result is proportional to $p^4$ and will only renormalize higher-derivative operators. Additionally, we see that the dependence on the gauge parameter $z$, which appears explicitly in the ghost propagators (\ref{eq:prop_ghost1}-\ref{eq:prop_ghost3}), has cancelled out in the final result. This cancellation is non-trivial, since individual diagrams depend on $z$ and only the total combination is independent of the parameter.

\subsubsection{The total result}
We finally add up all the different contributions computed in the previous sections, finding that the total one-loop correction to the two-point function of the background graviton is
\begin{align}\label{eq:2-pt_graviton}
\nonumber &\langle H_{\m\n}(-p) H_{\a\b}(p)\rangle_{\rm 1-loop}=\langle H_{\m\n}(-p) H_{\a\b}(p)\rangle_{\phi}+\langle H_{\m\n}(-p) H_{\a\b}(p)\rangle_{\rm bosons}+\langle H_{\m\n}(-p) H_{\a\b}(p)\rangle_{\rm fermions}\\
\nonumber &=\left[\frac{p^4 (50+100\sigma + 811\sigma^2) (\eta_{\a\m}\eta_{\b\n}+\eta_{\a\n}\eta_{\b\m})}{480 \sigma^2} + \frac{p^4 (-125-250\sigma +\sigma^2 (-2281 +80 \xi +240 \xi^2)) \eta_{\m\n}\eta_{\a\b}}{1920 \sigma^2}\right. \\
\nonumber & + \frac{p^2 (25+50\sigma +\sigma^2 (659-80\xi -240\xi^2) (\eta_{\m\n}p_\a p_\b +\eta_{\a\b} p_\m p_\n)}{480 \sigma^2} + \frac{9 (25 + 50 \sigma + 8 \sigma^2 (19 + 10 \xi  + 30 \xi ^2))p_\a p_\b p_\m p_\n}{1080 \sigma^2} \\
\nonumber & - \frac{p^2 (50+100\sigma +811\sigma^2)(p_\m p_\a \eta_{\n\b} + p_\n p_\a \eta_{\m\b} + p_\m p_\b \eta_{\n\a}+p_\n p_\b \eta_{\a\m})}{480\sigma^2}-\frac{m^2(1+6\xi)}{48}\left(4 p^2 (\eta_{\a\n} \eta_{\b\m} +\eta_{\a\m} \eta_{\b\n})\right.\\
&\left.\left.   + 4 (\eta_{\m\n} p_{\a} p_{\b} +\eta_{\a\b} p_{\m} p_{\n}) -4( \eta_{\b\n} p_{\a} p_{\m}+ \eta_{\a\n}p_{\b} p_{\m} + \eta_{\b\m} p_{\a} p_{\n} +\eta_{\a\m} p_{\b} p_{\n})-3 p^2 \eta_{\a\b} \eta_{\m\n}\right)\right]{\cal I}.
\end{align}

\subsection{Renormalization}
Once we have computed the divergent parts of the different correlation functions, we come to the moment of renormalizing the effective action, absorbing the divergences by using a counter-term. For any generic correlation function ${\cal G}$, we will compute the value ${\cal I}$ by using \eqref{eq:I_integral} so that we will have
\begin{align}
{\cal G}_{\rm 1-loop}\equiv \bar {\cal G} \left(\frac{i}{8\pi^2 \epsilon}-\frac{i}{16\pi^2}\left(\gamma-\log(4\pi)+\log(\eta^2)\right)+{\cal O}(\epsilon)\right),
\end{align}
where $\bar{\cal G}$ will be a tensor structure depending on $p^\m$ and on the coupling constants of the theory. We will then add counterterms to the bare Lagrangian, including also higher-derivative new operators that we will need to absorb the divergences quartic in $p^\m$. Using the $\overline{MS}$ subtraction scheme then we write
\begin{align}
{\cal G}_{\rm ct}\propto \delta c  \left(\frac{i}{8\pi^2 \epsilon}-\frac{i}{16\pi^2}\left(\gamma-\log(4\pi)+\log(\mu^2)\right)\right)=\delta c \ \Re(\mu),
\end{align}
for a generic coupling $c$. We have defined
\begin{align}
\Re(\mu)= \left(\frac{i}{8\pi^2 \epsilon}-\frac{i}{16\pi^2}\left(\gamma-\log(4\pi)+\log(\mu^2)\right)\right),
\end{align}
where $\mu$ is the renormalization scale.

We will determine the value of $\delta c$ so that the the sum ${\cal G}_{\rm 1-loop}+{\cal G}_{\rm counter-term}$ is free of divergences when $\epsilon\rightarrow 0$.

\subsubsection{Scalar two-point function}
In order to absorb the divergences in the two-point function of the scalar field, we must extend the bare action by including an operator with four derivatives in the kinetic term. The corresponding action for the counter-terms can be written in the frame where the metric is unimodular in the standard way
\begin{align}\label{eq:CT2}
\delta S_{2,\phi}=\int d^4x \left( \frac{\delta Z}{2}\partial \phi \partial^\m \phi -\frac{\delta m^2}{2}\phi^2 +\frac{\delta a_4}{2} \square \phi  \square \phi\right),
\end{align}
where $a_4$ is a dimensionful coupling and $\delta Z$ is the anomalous dimension of the scalar field, related to the field strength renormalization as usual
\begin{align}
\phi_R= Z^{\frac{1}{2}}\phi,\quad Z=1+\delta Z.
\end{align}

Now, we perform the change of variables to the unconstrained background metric by
\begin{align}\label{eq:change_bg}
g_{\m\n}=|\bar g|^{\frac{1}{4}}\bar g_{\m\n},
\end{align}
for which the action takes the slightly more involved form
\begin{align}
\delta S_{2,\phi}=\int d^4x\ \left( \frac{\delta Z}{2}|\bar{g}|^{\frac{1}{4}}\partial_\m \phi \partial^\m \phi -\frac{\delta m^2}{2}\phi^2 +\frac{\delta a_4}{2}  | \bar g|^{\frac{1}{2}} D^2 \phi  D^2 \phi\right),
\end{align}
which is explicitly \emph{WTDiff} invariant.

The contribution from the counter-terms to the correlation function is then
\begin{align}
\langle \phi(-p)\phi(p)\rangle_{\rm ct}=i\left(\delta Z p^2 -\delta m^2 + \delta a_4 p^4\right)\Re(\mu).
\end{align}

Adding it to the one-loop result \eqref{eq:2pt_scalar} and demanding the result to be finite, we find that the value of the counter-terms must be
\begin{align}
&\delta Z=\frac{3 G m^2 (1 + 2 \sigma (1 + 4 \xi  (2 + \xi )))}{4 \sigma},\\
&\delta a_4=- \frac{G (3 + 2 \sigma (3 + 8 \xi ))}{4 \sigma},\\
&\delta m^2=12 \lambda m^2  -6 G m^4 \xi^2 .
\end{align}

\subsubsection{Scalar four-point function}
In order to renormalize the divergences in the four-point function \eqref{eq:4-pt_scalar} we also need to include higher-derivative operators. As before, we write them in a standard form in the unimodular frame
\begin{align}\label{eq:CT4}
S_{4,\phi}=\int d^4x\ \left( -\delta \lambda \phi^4+\frac{\delta b_2}{8}\phi^2 (\partial \phi)^2 +\frac{\delta b_4}{24}( \partial \phi)^4\right).
\end{align}

Writing it in the unconstrained frame with \eqref{eq:change_bg}, the corresponding action, which is invariant under background \emph{WTDiff} transformations, then reads
\begin{align}
S_{4,\phi}=\int d^4x\ \left( -\delta \lambda \phi^4+\frac{\delta b_2}{8}|g|^{\frac{1}{4}}\phi^2 (\partial \phi)^2 +\frac{\delta b_4}{24}|\bar{g}|^{\frac{1}{2}}(\partial \phi)^4\right),
\end{align}
and gives a contribution to the correlator of the form
\begin{align}
\langle \phi(-p)\phi(-p)\phi(p)\phi(p)\rangle_{\rm ct}=i\left(-24 \delta\lambda + \delta b_2 p^2 +\delta b_4 p^4\right)\Re(\mu).
\end{align}

Adding it to \eqref{eq:4-pt_scalar} and cancelling the divergences in $\epsilon$ we find
\begin{align}
&\delta \lambda=6 \lambda^2  -72 G \lambda m^2 \xi ^2 + \frac{3}{2} G^2 m^4 \xi ^3 (2 + 9 \xi ),\\
&\delta b_2 =- \frac{6 G (-6 \lambda (1 + \sigma (2 + 8 \xi  (4 + \xi ))) + G
m^2 \xi  (-3 \xi  + \sigma (1 + 2 \xi  (11 + 6 \xi  (7 + 3 \xi )))))}{\sigma},\\
&\delta b_4=- \frac{G^2 (117 + 8 \xi  (17 + 40 \xi )  -4 \sigma (27 + 4 \xi  (37 + 45 \xi )) + 4 \sigma^2 (483 + 4 \xi  (259 + 2 \xi  (385 + 3 \xi  (-28 + 9 \xi )))))}{8 \sigma^2},
\end{align}
so that the total correlation function in the one-loop approximation is now finite.

\subsubsection{The non-minimal coupling}
Now we come to the renormalization of the corrections to the non-minimal coupling, given by $\eqref{eq:non-minimal_one-loop}$. In order to do that we will need not only to introduce a counter-term for $\xi$ and a new higher-derivative operator, but also take into account the contribution of two operators that we have already included in a previous section, since they contain the metric and therefore will also contribute to this correlator when expanded around flat space. The full counter-term action that we need is then
\begin{align}\label{eq:CTnm}
S_{\phi\phi H}=\int d^4x \ \left( \frac{\delta Z}{2}\partial_\m \phi \partial^\m \phi+\frac{\delta a_4}{2}\square^2\phi \square^2\phi -\frac{\delta \xi}{2} \phi^2 R +\frac{\delta \varsigma}{2} \partial_\m\phi \partial^\mu \phi R\right),
\end{align}
which in the unconstrained frame reads
\begin{align}
\nonumber S_{\phi\phi H}=\int d^4x \ |\bar{g}|^{\frac{1}{4}}&\left[ \frac{\delta Z}{2} \partial_\m \phi \partial^\m \phi+\frac{\delta a_4}{2}|\bar g|^{\frac{1}{4}}D^2\phi D^2\phi -\frac{\delta \xi}{2} \phi^2 \left(\bar R + \frac{3 \bar 
\square |\bar{g}|}{4|\bar g|} - \frac{27 \bar \nabla_{\m} |\bar g| \bar \nabla^{\m}|\bar g|}{32 |\bar g|^2}\right)\right.\\
&\left.  +\frac{\delta \varsigma}{2}|\bar g|^{\frac{1}{4}} \partial_\m\phi \partial^\mu \phi \ \left(\bar R + \frac{3 \bar 
\square |\bar{g}|}{4|\bar g|} - \frac{27 \bar \nabla_{\m} |\bar g| \bar \nabla^{\m}|\bar g|}{32 |\bar g|^2}\right)\right].
\end{align}

The contribution from this action to the corresponding correlation function is then
\begin{align}
\nonumber \langle \phi(-p)\phi(-p)H_{\m\n}(2p)\rangle_{\rm ct}&=-i\left(\frac{p^2 (2\delta \varsigma + \delta a_4)}{2}+\frac{\delta Z +4 \delta \xi}{4}\right)\left(\eta_{\m\n}p^2 -4 p_\m p_\n\right)\Re(\mu)\\
&\equiv -i\left(p^2 \delta \varsigma+ \delta \xi\right)\left(\eta_{\m\n}p^2 -4 p_\m p_\n\right)\Re(\mu),
\end{align}
where in the last step we have absorbed the value of $\delta a_4$ and $\delta Z$ into the arbitrariness of $\delta \varsigma$ and $\delta \xi$ by redefining them.

Adding this to the divergent result \eqref{eq:non-minimal_one-loop} and cancelling the divergences we have
\begin{align}\label{eq:ct_non-minimal}
&\delta \xi=\frac{64 \lambda \sigma (1 + 6 \xi ) + G m^2 (3 + \sigma (6 + 8 \xi  (6 + (5  -24 \xi ) \xi )))}{16 \sigma},\\
&\delta  \varsigma=\frac{G (9 + 40 \xi  + 2 \sigma (-23  -30 \xi  + 2 \sigma (-21 + 2 \xi  (-53 + 6 \xi  (-13 + 6 \xi )))))}{48 \sigma^2}.
\end{align}
\subsubsection{Gravitational two-point function}
The last correlation function that we need to renormalize is the two-point function of the background graviton $\langle H_{\m\n}(-p) H_{\a\b}\rangle$. In order to absorb all the divergences we need to add counter-terms for the Newton's constant $G$ as well as two new standard higher-derivative operators in the form of $R^2$ and $R_{\m\n}R^{\m\n}$, that we write in the following combination
\begin{align}\label{eq:CTg2}
S_{2,H}=\int d^4x \left(-\delta\left(\frac{1}{2G}\right)R+\delta \alpha R^2 +\delta \rho\left(R_{\m\n}R^{\m\n}-\frac{1}{3}R^2\right)\right).
\end{align}

In principle we are allowed to add also a term $R_{\m\n\a\b}R^{\m\n\a\b}$. However its integral corresponds to the Gauss-Bonnet term in four space-time dimensions and therefore its variation -- and consequently its expansion around flat space -- vanishes.

Of course, the counter-terms must be now written in the unconstrained frame by performing the change of variables $g_{\m\n}=|\bar{g}|^{\frac{1}{4}}\bar{g}_{\m\n}$ for which we have
\begin{align}
&R_{\m\n}=\bar R_{\m\n} + \frac{\bar{g}_{\m\n} \partial_{\a}\partial^{\a}|\bar{g}|}{8 |\bar g|} - \frac{5 g_{\m\n} \partial_{\a}|\bar g| \partial^{\a}|\bar g|}{32 |\bar g|^2} + \frac{\partial_{\m}\partial_{\n}|\bar g|}{4 |\bar g|} - \frac{7 \partial_{\m}|\bar g| \partial_{\n}|\bar g|}{32 |\bar g|^2},\\
&R=\bar R + \frac{3  
\square |\bar{g}|}{4|\bar g|} - \frac{27 \partial_{\m} |\bar g|\partial^{\m}|\bar g|}{32 |\bar g|^2}.
\end{align}
We omit the full expression for $S_{2,H}$ since it is very cumbersome. Note that, since $G$ multiplies the kinetic term of the graviton, there is no need to introduce a field strength renormalization for $H_{\m\n}$.

The contribution of the counter-terms to the correlator is
\begin{align}
\nonumber \langle H_{\m\n}(-p) H_{\a\b}(p)\rangle_{\rm ct}=&\bigg\{-\frac{1}{4}\delta\left(\frac{1}{2G}\right)\bigg(-p^2(\eta_{\m\a}\eta_{\n\b}+\eta_{\m\b}\eta_{\n\a})+\frac{3p^2}{4}\eta_{\m\n}\eta_{\a\b}-(\eta_{\m\n}p_\a p_\b+\eta_{\a\b}p_\m p_\n) \\
\nonumber & +(\eta_{\m\a}p_\b p_\n + \eta_{\m\b}p_\a p_\n +\eta_{\n\a}p_\m p_\b +\eta_{\n\b}p_\m p_\a)\bigg)+\delta \alpha \bigg( \frac{p^4}{8} \eta_{\m\n}\eta_{\a\b}-\frac{p^2}{2}(\eta_{\m\n}p_\a p_\b +\eta_{\a\b}p_\m p_\n)\\
\nonumber &+ 2 p_\m p_\n p_\a p_\b\bigg)+\delta\rho \bigg(\frac{p^4}{4}(\eta_{\m\a}\eta_{\b\n} +\eta_{\m\b}\eta_{\n\a})-\frac{p^4}{6}\eta_{\m\n}\eta_{\a\b}+\frac{p^2}{6}(\eta_{\m\n}p_\a p_\b+\eta_{\a\b}p_\m p_\n)\\
&-\frac{p^2}{4}(\eta_{\m\a}p_\n p_\b +\eta_{\n\a}p_\m p_\b + \eta_{\m\b}p_\n p_a + \eta_{\n\b}p_\m p_\a)+\frac{p^2}{3}p_\a p_\b p_\m p_\n\bigg)\bigg\}\Re(\mu).
\end{align}

Adding this to \eqref{eq:2-pt_graviton} and demanding that the divergences cancel, we find
\begin{align}
&\delta\left(\frac{1}{2G}\right)=\frac{m^2(1+6\xi)}{3},\\
&\delta \alpha=-\frac{5 + 10 \sigma  - \sigma^2 (71  -48 \xi   -144 \xi ^2)}{144 \sigma^2},\\
&\delta \rho=-\frac{50 +100 \sigma + 811 \sigma^2}{120 \sigma^2}.
\end{align}

\section{$\beta$-functions and Running Couplings}\label{sec:beta_functions}
Once we have determined the form of the renormalized correlation functions that we need, we come to the issue of computing the renormalization group flow of the different coupling constants, which is independent of the renormalization and regularization schemes used in the previous sections. We will actually define the running of a given coupling through the Callan-Symanzik (CS) equation for the corresponding correlation function \cite{PhysRevD.2.1541,Symanzik:1970rt}
\begin{align}\label{eq:CS}
\left(\mu  \frac{\partial}{\partial \mu}+\sum_i \beta(c_i)\frac{\partial}{\partial c_i}+ \sum_j \gamma(M_j)M_j \frac{\partial}{\partial M_J}+n \gamma_\phi\right){\cal G}(p,\mu)=0,
\end{align}
which is obtained by demanding independence of the arbitrary scale $\mu$ introduced by renormalization. Here $c_i$ are all possible dimensionless couplings appearing in the correlator, while $M_j$ refers to dimensionful couplings. $\gamma(M_j)$ is then the \emph{anomalous dimension} of the coupling, while $\gamma_\phi$ is the anomalous dimension of the scalar field, being $n$ the number of external scalar legs in the correlator. Here we have already taken into account that the anomalous dimension of $H_{\m\n}$ vanishes. Solving the equation \eqref{eq:CS} perturbatively for the couplings in the action will allow us to obtain the running of all of them.

Since we are working at one-loop, we find an important simplification here. The only part of the renormalized correlation function which depends on $\mu$ is the counter-term, so we can make the replacement
\begin{align}
\mu \frac{\partial {\cal G}(p,\mu)}{\partial \mu}\equiv \mu \frac{\partial {\cal G}_{\rm ct}(p,\mu)}{\partial \mu}.
\end{align}

Additionally, since our expansion is polynomial in the couplings, derivatives with respect to them are ordered in the loop expansion, with increasing loops contributing with higher orders. For the one-loop computation at hand, this means that we can also replace
\begin{align}
\left(\sum_i \beta(c_i)\frac{\partial}{\partial c_i}+ \sum_j \gamma(M_j)M_j \frac{\partial}{\partial M_J}+n \gamma_\phi\right){\cal G}(p,\mu)\equiv \left(\sum_i \beta(c_i)\frac{\partial}{\partial c_i}+ \sum_j \gamma(M_j)M_j \frac{\partial}{\partial M_J}+n \gamma_\phi\right){\cal G}_{\rm tree}(p,\mu),
\end{align}  
since when acting on the corrections we will generate a next-to-leading-order term. These two substitutions simplify the computation greatly.

Let us then start by writing the simplified form of equation \eqref{eq:CS} for the two-point function of the scalar field
\begin{align}
\left(\mu \frac{\partial}{\partial \mu}+ m^2 \gamma(m^2)\frac{\partial}{\partial m^2}+a_4 \gamma(a_4) \frac{\partial}{\partial a_4}+2 \gamma_\phi\right)\langle \phi(-p)\phi(p)\rangle=0.
\end{align}

Combining the one-loop correction and the counter-term, and ordering this equation by powers of the momentum, it can be easily solved to get
\begin{align}
&\gamma_\phi=\frac{3 G m^2 (1 + 2 \sigma (1 + 4 \xi  (2 + \xi )))}{64 \pi^2 \sigma},\\
&\gamma(m^2)=\frac{3 \lambda}{2 \pi^2} - \frac{3 G m^2 (1 + 2 \sigma (1 + 8 \xi  (1 + \xi )))}{32 \pi^2 \sigma},\\
&\gamma(a_4)=-G\left( \frac{3 + 2 \sigma (3 + 8 \xi )}{32 a_4 \pi^2 \sigma}  + \frac{3 m^2 (1 + 2 \sigma (1 + 8\xi  + 4 \xi ^2))}{32 \pi^2 \sigma}\right).
\end{align}

From the four-point function of the scalar field we can now compute the running of $\lambda$, $b_2$ and $b_4$. The corresponding CS equation is
\begin{align}\label{eq:CS_4pt}
\left(\mu \frac{\partial}{\partial \mu}+\beta(\lambda)\frac{\partial}{\partial \lambda}+b_2 \gamma(b_2)\frac{\partial }{\partial b_2}+b_4 \gamma(b_4) \frac{\partial}{\partial b_4}+4 \gamma_\phi\right)\langle \phi (-p)\phi(-p)\phi(p)\phi(p)\rangle=0.
\end{align}

Note however that this imposes a limitation in our computation. While the divergent contribution \eqref{eq:4-pt_scalar} contains pieces proportional to $G^2$, the field strength of the scalar field is only linear in $G$. This means that we should expect two-loop contributions to $\gamma_\phi$ of order $G^2$. Indeed if one notes that the powers of $G$ are brought into the diagrams by gravitational propagators, we can straightforwardly see that the following two diagrams, for instance, will potentially contribute to $\gamma_s$ at order $G^2$
\begin{align}
\begin{fmffile}{tl1} 
\parbox{15mm}{
\begin{fmfgraph*}(80,80) 
\fmftop{t}
\fmfbottom{b}
\fmfleft{l}
\fmfright{r}
\fmf{plain}{l,li}
\fmf{plain}{ri,r}
\fmf{phantom}{t,ti}
\fmf{phantom}{bi,b}
\fmf{photon,left=.4,tension=0.4}{li,ti}
\fmf{plain,left=.4,tension=0.4}{ti,ri}
\fmf{photon,left=.4,tension=0.4}{ri,bi}
\fmf{plain,left=.4,tension=0.4}{bi,li}
\fmf{plain, tension=.1}{bi,ti}
\end{fmfgraph*}
}\end{fmffile}\qquad  \quad \qquad , \quad 
\begin{fmffile}{tl2} 
\parbox{15mm}{
\begin{fmfgraph*}(80,80) 
\fmftop{t}
\fmfbottom{b}
\fmfleft{l}
\fmfright{r}
\fmf{plain}{l,li}
\fmf{plain}{ri,r}
\fmf{photon,left=1,tension=.1}{li,ci}
\fmf{plain,right=1,tension=.1}{li,ci}
\fmf{photon,left=1,tension=.1}{ri,ci}
\fmf{plain,right=1,tension=.1}{ri,ci}
\end{fmfgraph*}
}\end{fmffile}\qquad \qquad .
\end{align}

Therefore, if we wanted to solve the CS equation \eqref{eq:CS_4pt} at order $G^2$ we would need to add the contribution coming from the two-loop correction to $\gamma_s$. As a consequence, we can only trust our result here up to order $G$ and thus we will cut the perturbative solution to \eqref{eq:CS_4pt} at this order. It reads
\begin{align}
&\beta(\lambda)=\frac{9 \lambda^2}{2 \pi^2}  - \frac{3 G \lambda m^2 (1 + 2\sigma (1 + 8 \xi  + 28 \xi ^2))}{16 \pi^2 \sigma},\\
&\gamma(b_2)=G\left(- \frac{3 m^2 (1 + 2 \sigma (1 + 8 \xi  + 4 \xi ^2))}{16 \pi^2 \sigma} + \frac{9 \lambda (1 +\sigma (2 + 32 \xi  + 8 \xi ^2))}{2 b_2 \pi^2 \sigma}\right),\\
&\gamma(b_4)=- \frac{3 G m^2 (1 + 2 \sigma (1 + 8 \xi  + 4 \xi ^2))}{16 \pi^2 \sigma}.
\end{align}

In the limit of decoupling gravitation $G\rightarrow 0$, the running of higher-derivative terms freeze as expected, while the running of $\lambda$ matches the text-book result\footnote{Note however that we are defining our coupling without the standard $4!$ denominator.} for $\lambda \phi^4$.

We will find the same issue previously discussed when trying to solve the CS equation for the running of the non-minimal coupling
\begin{align}
\left( \mu \frac{\partial}{\partial \mu}+\beta(\xi)\frac{\partial}{\partial \xi}+a_4 \gamma(a_4)\frac{\partial}{\partial a_4} + \varsigma \gamma(\varsigma)\frac{\partial}{\partial \varsigma}+2 \gamma_\phi\right)\langle \phi(-p)\phi(-p)H_{\m\n}(2p)\rangle=0,
\end{align}
since both $\gamma(a_4)$ and $\gamma_\phi$ are of order $G$ and we expect corrections of order $G^2$ coming from two-loop divergences. 

We therefore again cut the solution to this equation at order $G$. Taking the form of the divergence \eqref{eq:non-minimal_one-loop} and the counter-terms \eqref{eq:ct_non-minimal} and ordering the CS equation in powers of $p^\mu$ we find
\begin{align}
&\beta(\xi)=\frac{\lambda (1 + 6 \xi )}{2 \pi^2}  - \frac{G m^2 \xi  (3 + \sigma (6 + 44 \xi  + 72 \xi ^2))}{32 \pi^2 \sigma},\\
&\gamma(\varsigma)=\frac{G (9 + 40 \xi   -4 \sigma (7 + 15 \xi  + 9 m^2\varsigma ) + 8 \sigma^2 (-6  -41 \xi   -78 \xi ^2 + 36 \xi ^3  -9 m^2 (1 + 4 \xi  (2 + \xi )) \varsigma ))}{384 \pi^2 \sigma^2 \varsigma }.
\end{align}

Finally, we write the corresponding equation for the two-point function of the graviton
\begin{align}
\left( \mu \frac{\partial}{\partial \mu}+ G\gamma(G) \frac{\partial}{\partial
 G} +\beta(\alpha)\frac{\partial}{\partial \alpha}+ \beta(\rho)\frac{\partial}{\partial \rho}\right)\langle H_{\m\n}(-p)H_{\a\b}(p)\rangle=0.
\end{align}

Ordering it by powers of the momentum, we complete the computation of the renormalization group flow of the theory with the results
\begin{align}
&\gamma(G)=-\frac{Gm^2 (1+6\xi)}{12 \pi^2},\\
&\beta(\alpha)=-\frac{5 + 10 \sigma  - \sigma^2 (71  -48 \xi   -144 \xi ^2)}{1152 \pi^2 \sigma^2},\\
&\beta(\rho)=- \frac{50+ 100 \sigma + 811 \sigma^2}{960 \pi^2 \sigma^2}.
\end{align}

As a final note in this section, let us note that all the $\beta$ functions and $\gamma$-functions defined here reduce to those of the case of background gravity once the right limit $G\rightarrow 0$ is taken. The only subtlety comes from the running of $\alpha$ and $\beta$, which are however irrelevant because in the limit $G\rightarrow 0$ they are subleading with respect to the Einstein-Hilbert term in the action. 

This completes the computation of the one-loop $\beta$-functions for the unimodular scalar tensor-theory. In the appendix \ref{app:UG_vacuum} we discuss the results in the absence of the scalar field, in order to establish a comparison with previous works.

\section{Unimodular Gravity Versus General Relativity}\label{sec:versus}
Once the renormalization group flow of the coupling constants is computed, we come to the question that originated this work -- is Unimodular Gravity equivalent to General Relativity when coupled to matter? 

Although the question is simple enough, the answer is not so. First of all, we note that although the quantization of UG looks much more complicated than the one of GR, no new counter-terms are required in order to absorb all one-loop divergences. Indeed, all required counter-terms -- depicted in \eqref{eq:CT2}, \eqref{eq:CT4}, \eqref{eq:CTnm}, and \eqref{eq:CTg2}-- are exactly the same ones that would be required in GR, just appended with the condition $|g|=1$.

In order to differentiate both theories we shall then look at a physical observable. However, the running of the couplings that we have just derived does not classify as such, as it can be observed by the explicit dependence on the gauge fixing parameter $\sigma$ of most of them. Moreover, some of our results could, in principle, be modified by a non-linear redefinition of the gravitational field $H_{\m\n}\rightarrow H_{\m\n}(\phi)$, clearly denoting that they lack a physical meaning due to operator mixing after the field redefinition \cite{Gonzalez-Martin:2017bvw,Anber:2010uj}. To determine something which can be thought as physical, we must then find which combinations of the couplings are independent of the gauge choice and blind to field redefinitions. Those, known as \emph{essential couplings}, will be the couplings that control correlation functions of observable quantities. Only the $\beta$-functions of essential couplings have an intrinsic physical meaning. They can be determined by noting that they correspond to the only combinations that do not change when we add to the action a piece proportional to the classical equations of motion \cite{DeWitt:1967ub,Kallosh:1974yh}.

In the following we will focus only on those couplings which are present in the bare Lagrangian, ignoring the higher-derivative operators. Moreover, and for simplicity, we will consider solutions to the eom with unimodular background determinant ${|\bar{g}|}=1$. Under this assumption, the background action reads
\begin{align}
S=\int d^4x\bigg(& -\frac{1}{2G} \bar R  + \frac{1}{2}\partial_\m \phi \partial^\m \phi -\frac{\xi}{2}\phi^2 \bar R-\frac{m^2}{2}\phi^2 -\lambda \phi^4\bigg).
\end{align}

In this frame, the equations of motion are the traceless Einstein equations \eqref{eq:einstein_traceless}, which by using Bianchi identities are equivalent to the full set of Einstein equations with an arbitrary cosmological constant ${\cal C}$
\begin{align}
\bar R_{\m\n}-\frac{1}{2}\bar R g_{\m\n} +{\cal C}\bar g_{\m\n}=G T_{\m\n}.
\end{align}

The only scalar quantity up to two derivatives that we can form with the eom is then the trace of Einstein equations
\begin{align}
{\cal E}=\bar R+GT-4{\cal C}=\bar R+G\left[-(1+6\xi) \partial_\m \phi \partial^\m \phi -6\xi \phi \square \phi +2  m^2 \phi^2 +4 \lambda \phi^4 +\xi G \phi^2 \bar R\right] -4 {\cal C},
\end{align}
where we have used
\begin{align}
T_{\m\n}=(1+2\xi)\partial_\m \phi \partial_\n\phi +2\xi \phi \bar \nabla_\m \partial_\n \phi -\xi \phi^2\bar{R}_{\m\n}-\bar g_{\m\n}\left(\frac{1+4\xi}{2}\partial_\a \phi \partial^\a \phi+2\xi \phi \square \phi-\frac{m^2}{2}\phi^2-\lambda \phi^4 -\frac{\xi}{2}\phi^2 \bar R\right). 
\end{align}

We thus add a piece proportional to the trace of the eom to the action
\begin{align}\label{eq:shift_action}
S\rightarrow S+\int d^4x \  \frac{\Sigma {\cal E}}{2G},
\end{align}
where $\Sigma$ is a constant parameter. Under this addition, ignoring the cosmological constant and integrating by parts, we find that the couplings transform as
\begin{align}
&\delta_\Sigma G=\Sigma G,\\
&\delta_\Sigma m^2=-2\Sigma m^2 ,\\
&\delta_\Sigma \lambda=-2\Sigma \lambda,\\
&\delta_\Sigma \xi=-\Sigma \xi.
\end{align}

Thus, essential couplings will be combinations of these that are invariant under the addition of the evanescent piece proportional to $\Sigma$. Additionally, since we want to avoid the arbitrariness tied to the reference scale for dimensionful quantities, we will demand our essential couplings to be dimensionless as well.

Out of $G$ and $m^2$ we can build the following scale-invariant coupling
\begin{align}
\mathfrak{G}=Gm^2,\qquad \delta \mathfrak{G}=- \Sigma \mathfrak{G},
\end{align}
which transforms as $\sqrt{\lambda}$ and therefore we can build a ratio which is an essential coupling, given by
\begin{align}
\Delta =\frac{\mathfrak{G}^2}{\lambda}.
\end{align}
It corresponds to the relative strength between interactions dictated by the Einstein-Hilbert term and the self-interaction of the scalar in the Lagrangian, as measured in a $2\rightarrow 2$ scattering of scalar fields. The two channels involved in this process, graviton exchange and contact interactions, are schematically given by
\begin{align}
\begin{fmffile}{exchange} 
\parbox{15mm}{
\begin{fmfgraph*}(70,50) 
\fmfleft{l1,l2}
\fmfright{r1,r2}
\fmf{plain}{l1,li}
\fmf{plain}{l2,li}
\fmf{plain}{ri,r2}
\fmf{plain}{ri,r1}
\fmf{photon}{li,ri}
\end{fmfgraph*}
}\end{fmffile}\quad\qquad  \sim \frac{G m^4}{p^2}=\mathfrak{G}^2 \left(G p^2\right)^{-1},\qquad \begin{fmffile}{contact} 
\parbox{15mm}{
\begin{fmfgraph*}(70,50) 
\fmfleft{l1,l2}
\fmfright{r1,r2}
\fmf{plain}{l1,li}
\fmf{plain}{l2,li}
\fmf{plain}{li,r2}
\fmf{plain}{li,r1}
\end{fmfgraph*}
}\end{fmffile}\quad \qquad \sim \lambda,
\end{align}
and we can see that their ratio corresponds to $\Delta$ when the momentum exchanged by the graviton is set at the scale where gravitational interactions dominate $p^2\sim G^{-1}\sim M_P^2$. Thus, the running of $\Delta$ indicates at which energy scale gravitation becomes important and cannot be ignored when doing QFT with scalar fields. Its $\beta$-function can be easily obtained from the ones of the couplings in the Lagrangian. It reads
\begin{align}\label{eq:Delta}
&\beta(\Delta)=\frac{\Delta(-9 \lambda + \mathfrak{G} (-1  -6 \xi  + 45\xi^2))}{6  \pi^2},
\end{align}
where as expected, the dependence on the gauge parameter $\sigma$ has cancelled out. In principle, we could also define two other ratios $\Delta_i$ involving $\xi$. However, these enter strong-coupling when either $\mathfrak{G},\lambda,\xi\rightarrow 0$ since their $\beta$-functions are not polynomial in the couplings\footnote{This can be seen by writing the $\beta$-functions in the form
\begin{align*}
\beta(\Delta_i)=\Delta_i(\dots),
\end{align*}
where the dots indicate an expansion in the couplings of the theory, which will depend on the particular coupling chosen. For instance, for $\Delta_\xi=\mathfrak{G}/\xi$ the leading term within the parenthesis goes as $\xi^{-1}$ and therefore it is not perturbative in the sense discussed along this work. Note that this is not the case for the coupling \eqref{eq:Delta}.}. We will therefore refrain from discussing them hereinafter.

An unpleasant property about $\beta(\Delta)$ that we must remark here is that although $\Delta$ is an essential coupling, its running, albeit being gauge invariant, depends on the non-physical quantities $\lambda,\mathfrak{G}$ and $\xi$ independently. A similar property has been already noted before in the context of asymptotic safety\footnote{We are grateful to R. Percacci for pointing this out.} for the running of essential couplings \cite{Dou:1997fg,Benedetti:2011ct}. It implies that in this situation one cannot disentangle physical contributions from un-physical ones but instead one needs to first compute the latter in order to derive the former. It also poses a conundrum on understanding how the value of $\Delta$ can indeed remain essential along the RG flow and at higher order in perturbation theory. Here we cannot offer any satisfactory explanation beyond hinting that this might be a consequence of the non-renormalizability of the theory.

Following the same reasoning depicted before, we see that the same definition of essential couplings holds for the case of GR, when we restore the $\sqrt{|g|}$ in the integration measure and shift the action accordingly by modifying \eqref{eq:shift_action}. Although the value for the running of the couplings in GR can be found in the literature \cite{Steinwachs:2011zs}, we prefer here to re-derive the needed ones with the same techniques described for UG. Details of this computation can be found in appendix \ref{app:GR}. In that case, the value of the $\beta$-function for the essential coupling takes the form
\begin{align}\label{eq:betaGR}
&\beta_{\rm GR}(\Delta)=\frac{\Delta(-9 \lambda + \mathfrak{G} (-1 +39 \xi  +45 \xi ^2))}{6  \pi^2},
\end{align}
which is subtlety but clearly different from \eqref{eq:Delta}.

Let us remark that the coupling $\Delta$ has a physical meaning. It gives a definite answer to the question of when gravitational interactions can be disregarded. As such, the fact that it does not agree with the UG result is a smoking gun that the theories cannot be considered equivalent once gravitation is dynamical. However, we see that the difference is very minor. The one-loop result $\beta(\Delta)$ agrees in both theories in two very important limiting cases, $\xi\rightarrow 0$ and $\xi \gg 1$.

The first limit  corresponds to a scalar field minimally coupled. In that case, we see that although the theory is very complicated, the running of the physical parameter $\Delta$ is identical to the more easily computed one in GR. It seems that non-minimal coupling is then an important ingredient to violate the equivalence. One could argue then that the full identification of both theories seems to be connected in a very non-trivial way to the satisfaction of the strong equivalence principle.

The second case is also interesting, since it corresponds to the limit in which several models of inflation -- in particular Higgs \cite{Bezrukov:2007ep,Rubio:2018ogq} and Higg-Dilaton inflation \cite{Bezrukov:2012hx,GarciaBellido:2011de,Shaposhnikov:2008xi} -- are successful. Although strictly speaking we have performed our computations in the limit $\xi \ll 1$ and therefore they would not be valid in the large $\xi$ limit, let us note that in the case $\xi\gg 1$ we can also take $G \gg 1$ and then the role of both couplings is formally exchanged in the action for the case of approximately constant scalar profiles. In that case, the equations for gravity reduce to
\begin{align}
\xi \phi^2 \left(R_{\m\n}-\frac{1}{4}R g_{\m\n}+{\cal O}\left(\frac{1}{\xi}\right)\right)=0,
\end{align}
which corresponds to the vacuum equations, up to sub-leading corrections. This means that in the $\xi \gg 1$ limit, and around flat space, the theory becomes indistinguishable from the case $\xi\ll 1$ and therefore our result should hold.

In any other intermediate value of $\xi$ we find that UG and GR \emph{are not equivalent}. Although this might look minor, since the difference is very subtle, it might have influence in intermediate energy regimes when moving along the flow. For instance, in the thermal history of our Universe. It also poses a question mark on the validity of quantum computations performed in UG without taking into account the very complicated quantization structure and just assuming that, since they are classically equivalent, one can compute in GR instead. This is clearly wrong at the light of our result.

Let us finally remark that we strongly believe in the robustness of our result. We have derived it independently by using two different computer codes in different languages. Moreover, the preservation of gauge invariance -- both at the background level and from independence of $\sigma$ -- is a very non-trivial issue and any minor modification of any ingredient in the computation would produce a result not satisfying it.

\section{Discussion and Conclusions}\label{sec:conclusions}
In this paper we have studied the question of the equivalence between General Relativity and Unimodular Gravity. Although the answer is positive when we look to classical physics, or even classical gravitation in the presence of quantum matter, there are important subtleties when gravitons are dynamical and allowed to run freely in the loops.

In order to discuss this property in a QFT manner, we started by formulating the theory in a frame where the constraint $|g|=1$ is automatically satisfied, by redefining the metric as $g_{\m\n}=|\tilde{g}|^{\frac{1}{4}} \tilde{g}_{\m\n}$. In this frame UG becomes a pretty non-standard theory enjoying an extended gauge symmetry, the product of \emph{TDiff} and \emph{Weyl}, that we call \emph{WTDiff}. However, in this form the main properties of UG are explicit. The counting of degrees of freedom is straightfoward and the traceless character of the eom is explicit. In order to compute one-loop corrections we exploited this symmetry by using the background field method in combination with the construction of a Weyl invariant geometry. 

The construction of the gauge sector -- combining gauge fixing and ghost action -- becomes surprisingly much more cumbersome in UG than in GR, mainly due to the fact that \emph{TDiff} generators are not independent but rather constrained to be a transverse vector. Although they can then be represented by using a transverse projector and the full gauge system solved by using BRST symmetry, this generates a tower of new ghost fields of bosonic character. These fields actually couple with the graviton degree of freedom, showing a first non-trivial difference of UG with respect to GR, at least at the technical level. Even for tree-level computations, one cannot just ignore the gauge sector, since the kinetic mixing between $h_{\m\n}$ and the bosonic ghosts will have an impact on the propagator of the former.

Nevertheless, once the issue with constructing the gauge sector is solved, then the one-loop corrections to the correlation functions of the theory can be computed in a standard manner by expanding the background around flat space and looking at Feynman diagrams carrying perturbations of the background in the external legs. Although there are plenty of them -- in particular due to the mixing of the graviton with the bosonic fields --, this is a task that we were able to carry out with the help of computer codes specialized in tensor algebra.

The result of our computations are the complete set of $\beta$-functions and anomalous dimensions of all the couplings involved in the action, computed in the one-loop approximation and at order $G$, which we reproduce here to collect them together
\begin{align}
\nonumber &\gamma_\phi=\frac{3 G m^2 (1 + 2 \sigma (1 + 4 \xi  (2 + \xi )))}{64 \pi^2 \sigma},\\
\nonumber&\gamma(m^2)=\frac{3 \lambda}{2 \pi^2} - \frac{3 G m^2 (1 + 2 \sigma (1 + 8 \xi  (1 + \xi )))}{32 \pi^2 \sigma},\\
\nonumber&\gamma(a_4)=-G\left( \frac{3 + 2 \sigma (3 + 8 \xi )}{32 a_4 \pi^2 \sigma}  + \frac{3 m^2 (1 + 2 \sigma (1 + 8\xi  + 4 \xi ^2))}{32 \pi^2 \sigma}\right),\\
\nonumber&\beta(\lambda)=\frac{9 \lambda^2}{2 \pi^2}  - \frac{3 G \lambda m^2 (1 + 2\sigma (1 + 8 \xi  + 28 \xi ^2))}{16 \pi^2 \sigma},\\
\nonumber&\gamma(b_2)=G\left(- \frac{3 m^2 (1 + 2 \sigma (1 + 8 \xi  + 4 \xi ^2))}{16 \pi^2 \sigma} + \frac{9 \lambda (1 +\sigma (2 + 32 \xi  + 8 \xi ^2))}{2 b_2 \pi^2 \sigma}\right),\\
\nonumber&\gamma(b_4)=- \frac{3 G m^2 (1 + 2 \sigma (1 + 8 \xi  + 4 \xi ^2))}{16 \pi^2 \sigma},\\
\nonumber&\beta(\xi)=\frac{\lambda (1 + 6 \xi )}{2 \pi^2}  - \frac{G m^2 \xi  (3 + \sigma (6 + 44 \xi  + 72 \xi ^2))}{32 \pi^2 \sigma},\\
\nonumber&\gamma(\varsigma)=\frac{G (9 + 40 \xi   -4 \sigma (7 + 15 \xi  + 9 m^2\varsigma ) + 8 \sigma^2 (-6  -41 \xi   -78 \xi ^2 + 36 \xi ^3  -9 m^2 (1 + 4 \xi  (2 + \xi )) \varsigma ))}{384 \pi^2 \sigma^2 \varsigma },\\
\nonumber &\gamma(G)=-\frac{Gm^2 (1+6\xi)}{12 \pi^2},\\
\nonumber &\beta(\alpha)=-\frac{5 + 10 \sigma  - \sigma^2 (71  -48 \xi   -144 \xi ^2)}{1152 \pi^2 \sigma^2},\\
&\beta(\rho)=- \frac{50+ 100 \sigma + 811 \sigma^2}{960 \pi^2 \sigma^2}.
\end{align}
They include the couplings present in the classical Lagrangian but also new couplings controlling the strength of higher-derivative terms in the EFT expansion, as required by the non-renormalizability of gravity. The full one-loop EFT action that we obtain is then
\begin{align}
\nonumber S_{\rm 1-loop}=\int d^4x\bigg[ &-\frac{1}{2G}  R  + \frac{1}{2}\partial_\m \phi \partial^\m \phi -\frac{\xi}{2}\phi^2  R-\frac{m^2}{2}\phi^2 -\lambda \phi^4+\frac{ a_4}{2} \square\phi  \square \phi+\frac{ b_2}{8}\phi^2 (\partial \phi)^2 \\
&+\frac{ b_4}{24}( \partial \phi)^4+\frac{\varsigma}{2} \partial_\m\phi \partial^\mu \phi R+ \alpha R^2 + \rho\left(R_{\m\n}R^{\m\n}-\frac{1}{3}R^2\right)\bigg],
\end{align}
in the frame where the metric is unimodular $|g|=1$.

These runnings are however dependent on the gauge choice used to quantize the theory and therefore they do not correspond to physical quantities. Out of them, we identify the combination $\Delta=G^2 m^4 \lambda^{-1}$, which controls the relative strength of gravitational interactions with respect to scalar self-interactions and therefore has a physical meaning. It corresponds to an essential coupling of the theory. Its running is then gauge invariant and reads
\begin{align}
\beta(\Delta)=\frac{\Delta (-9 \lambda + \mathfrak{G} (-1  -6 \xi  + 45\xi^2))}{6\pi^2}.
\end{align}

We find that this quantity actually differs from the corresponding result in GR, which can be found in \eqref{eq:betaGR}, whenever the non-minimal coupling ranges on intermediate values. Only in the two extremal limits $\xi\rightarrow 0$ and $\xi \gg 1$, our result agrees with the general relativistic one. We interpret the first of these agreements as a consequence of the strong equivalence principle, which is then violated by non-minimal coupling. The second coincidence can be traced back to the singular behaviour of the eom in the large $\xi$ limit. However, the difference in intermediate regimes might be important when considering situations in which following the running of physical quantities along an energy history is critical, like the thermal history of the Universe.

Altogether this poses a question on the validity of several approaches found in the literature to computing quantum corrections in the case of UG. One cannot just assume that the theories are equivalent, restore $\sqrt{|g|}$ in the action, and compute quantum corrections in GR by hiding under the carpet the fact that actually one wants to work with UG. In particular, it would be interesting to revisit these results and their effects in models of inflation, which closely resemble the case studied here.
\section*{Acknowledgments}
We are grateful to E. Álvarez, C.P. Martin, S. Mooij, R. Percacci and C. Steinwachs for discussions and/or comments on the manuscript. M. H-V has been supported by the European Union's H2020 ERC Consolidator Grant “GRavity from Astrophysical to Microscopic Scales” grant agreement no. GRAMS-815673. R. S-G. has been supported by the Spanish FPU Grant No FPU16/01595 and by COST action CA16104 ``GWverse"  through a Short Term Scientific Mission. R. S-G. also wishes to thank the APP department at SISSA for their hospitality during part of this work. 

\appendix

\section{Computation of $\beta$-functions in General Relativity}\label{sec:GR} \label{app:GR}
We summarize here the computation of the $\beta$-function of the composite coupling $\Delta$ in GR, following the same techniques as in the case of UG. We will consider the action equivalent to $S_{\rm UG}+S_{\rm matter}$ by restoring $\sqrt{|g|}$
\begin{align}
S_{\rm GR}=\int d^4x\sqrt{|g|} \left(-\frac{1}{2G}R+\frac{1}{2}\partial_\m \phi \partial^\m \phi -\frac{m^2}{2}\phi^2 - \lambda\phi^4 -\frac{\xi}{2}\phi^2 R \right).
\end{align}

We will also expand this around a background configuration for the gravitational field $g_{\m\n}=\bar{g}_{\m\n}+h_{\m\n}$. However, the absence of Weyl invariance and the independence of the generators of \emph{Diff} allow us to construct a standard gauge fixing \emph{à la Feynman}
\begin{align}
S_{gf}=-\frac{\sigma}{2G}\int d^4x \sqrt{|\bar g|}\ F_\m F^\m, 
\end{align}
with $F_\m$ analogous to \eqref{eq:gauge_condition}
\begin{align}
F_\m=\bar{\nabla}^\n h_{\m\n}-\frac{1}{2}\bar \nabla_\m h.
\end{align}
Since we only want to compute the running of $\lambda$, $G$ and $m^2$, we will not need to add the action for the ghost fields in this case.

Expanding now the background metric around flat space $\bar{g}_{\m\n}=\eta_{\m\n}+H_{\m\n}$ and computing correlation functions involving $H_{\m\n}$ and $\phi$, we can derive the running of the couplings that we are interested in. The computation is analogous to that of UG with only two important changes -- the diagrams containing bosonic ghosts are now absent, and there are two extra diagrams to be considered
\begin{align}
\begin{fmffile}{extra_GR} 
\parbox{15mm}{
\begin{fmfgraph*}(80,80) 
\fmftop{t}
\fmfbottom{b}
\fmfleft{l1,l2,l3,l4,l5}
\fmfright{r}
\fmf{plain}{l2,li}
\fmf{plain}{l3,li}
\fmf{plain}{l4,li}
\fmf{plain}{ri,r}
\fmf{phantom}{t,ti}
\fmf{phantom}{bi,b}
\fmf{plain,left=.4,tension=0.6}{li,ti,ri}  
\fmf{photon,left=.4,tension=0.6}{ri,bi,li}  
\end{fmfgraph*}
}\end{fmffile}\qquad \qquad \qquad , \qquad 
\begin{fmffile}{extra2_GR} 
\parbox{15mm}{
\begin{fmfgraph*}(80,80) 
\fmftop{t}
\fmfbottom{b}
\fmfleft{l1,l2,l3,l4,l5}
\fmfright{r}
\fmf{plain}{l2,li}
\fmf{gluon}{l3,li}
\fmf{plain}{l4,li}
\fmf{phantom}{ri,r}
\fmf{phantom}{t,ti}
\fmf{phantom}{bi,b}
\fmf{plain,left=.4,tension=0.6}{li,ti,ri}  
\fmf{plain,left=.4,tension=0.6}{ri,bi,li}  
\end{fmfgraph*}
}\end{fmffile}\qquad ,
\end{align}
whose contributions are actually critical to ensure gauge invariance.

Running our code and computing all the Feynman diagrams, renormalizing, and solving the CS equation, we find
\begin{align}
&\gamma_{{\rm GR},\phi}=\frac{G m^2 (-1 + 6 \sigma (-1 - \xi  + \xi ^2))}{16 \pi^2 \sigma},\\
&\gamma_{\rm GR}(m^2)=\frac{6\lambda -3 G m^2 \xi  (1 + 2 \xi )}{4 \pi^2},\\
&\beta_{\rm GR}(\lambda)=\frac{3 \lambda (3 \lambda  - G m^2 \xi  (6 + 7 \xi ))}{2 \pi^2},\\
&\gamma_{\rm GR}(G)=-\frac{Gm^2 (1+6\xi)}{12 \pi^2},
\end{align}
from which we can compute the running of $\Delta=G m^2 \lambda^{-1}$ to be
\begin{align}
&\beta_{\rm GR}(\Delta)=\frac{\Delta (-9 \lambda + \mathfrak{G} (-1 +39 \xi  +45 \xi ^2))}{6 \pi^2},
\end{align}
where $\mathfrak{G}=G m^2$. We have also cross-checked the results of $\gamma_{\rm GR,\phi}$ and $\gamma_{\rm GR}(m^2)$ by using the three-point function mixing scalar fields and a graviton $\langle \phi(-p)\phi(-p)H_{\m\n}(2p)\rangle$.
\section{Quantum Corrections to Vacuum Unimodular Gravity}\label{sec:vacuum} \label{app:UG_vacuum}
For completeness, we take here a look to the renormalization group flow of the theory in the case of pure UG, when the action is only
\begin{align}
S_{\rm UG}=-\frac{1}{2G}\int d^4x \ |\tilde g|^{\frac{1}{4}}\ \left(\tR + \frac{3}{32}\frac{\tn_\m |\tg| \tn^\m |\tg|}{|\tg|^2}\right).
\end{align}

In this case, and using the renormalization scheme previously described in this work, we see that $G$ does not receive divergent one-loop corrections. Only the higher-derivative terms in \eqref{eq:CTg2} will run in this case. Subtracting the contribution of the scalar loops from our result and repeating the steps in section \ref{sec:computation}, we find that the running of the higher-derivative terms is controlled by
\begin{align}
&\beta_{\rm vacuum}(\alpha)=- \frac{5 + 10 \sigma  -30 \sigma^2}{1152 \pi^2 \sigma^2},\\
&\beta_{\rm vacuum}(\rho)=- \frac{50 + 100 \sigma + 807 \sigma^2}{960 \pi^2 \sigma^2}.
\end{align}

As we expected, these $\beta$-functions are gauge dependent, since they do not correspond to essential couplings. In order to check the robustness of our result we will then reconstruct the full non-linear divergence by noting that due to background field invariance, the divergent part of the one-loop correction -- the term proportional to $\epsilon^{-1}$ -- can be obtained by expanding the following action around flat space
\begin{align}\label{eq:full_CT}
S_{\rm div}=-\frac{i}{\epsilon}\int d^4x \left(\beta_{\rm vacuum}(\alpha)R^2 + \beta_{\rm vacuum}(\rho)\left(R_{\m\n}R^{\m\n}-\frac{1}{3}R^2\right)\right).
\end{align}
Otherwise, we would not be able to absorb it into a counter-term.

Now, gauge independence can be tested with the help of the Kallosh-DeWitt theorem \cite{DeWitt:1967ub,Kallosh:1974yh}. Since addition of terms proportional to the eom must be able to shift every gauge-dependent quantity, only when we take the previous divergence to be on-shell we must find a gauge-independent result. For the theory in vacuum, the eom of UG are equivalent to the full set of Einstein equations \eqref{eq:full_einstein}, which imply
\begin{align}
R_{\m\n}={\cal C} g_{\m\n}, \qquad R=4{\cal C}.
\end{align}

Plugging this into \eqref{eq:full_CT} we get
\begin{align}
S_{\rm div}=-\frac{173 i}{80 \pi^2 \epsilon}\int d^4x\ {\cal C}^2,
\end{align}
where the gauge dependence has vanished. Moreover, we find that the divergence is independent of the field, since there is no $\sqrt{| g|}$ term due to the unimodular condition. Therefore we conclude, in the same lines as \cite{Alvarez:2015sba}, that UG is one-loop finite \emph{even in the presence of a cosmological constant}.

Note however that our result cannot shed any light on the discrepancy of the results between \cite{Alvarez:2015sba} and \cite{Ardon:2017atk}, since here we only have access to gauge dependent quantities whose on-shell value is not dynamical. Due to the fact that we are working in perturbation theory around flat space, we cannot obtain the value of the topological term, which should be gauge independent by itself.

\bibliography{unimodular}{}

\end{document}